\documentclass[12pt]{article}

\usepackage{amssymb}

\usepackage{amsmath}

\usepackage{graphicx}
\usepackage{longtable}

\usepackage{bbm}
\usepackage{bm}
\usepackage{calc}
\usepackage{booktabs}
\usepackage{latexsym}
\usepackage{subcaption}
\usepackage{enumitem}   
\usepackage{float}

\newcommand{\sign}{\operatorname{sign}}
\newcommand{\Sign}{\operatorname{Sign}}

\newenvironment{keywords}{
	\vspace{1em}
	\noindent\textbf{Keywords:}
}{}

\begin{document}

\title{The effect of friction on the dynamics of targeted energy transfer by symmetric vibro-impact dampers}

\author{B.~Youssef, A.~Y.\ Karoui, R.~I.\ Leine}

\date{}
\maketitle
\begin{abstract}
This study investigates the nonlinear dynamics of a symmetric vibro-impact nonlinear energy sink (VI-NES) subjected to dry friction, a crucial factor that remains insufficiently explored in previous research. The combined effect of impact and friction leads to intricate behaviors that require further investigation. To address this, the multiple scales method is extended to incorporate frictional effects and is complemented with a generalized impact map approach. This allows for a systematic exploration of periodic solutions, stability, and bifurcations, revealing critical transitions between impact-dominated and sliding-dominated regimes. The activation thresholds and amplitude levels for different response regimes, including stick-slip dynamics, are identified, offering new insights into friction-induced nonlinearities. The results bridge the gap between theoretical modeling and practical implementation, offering a more accurate predictive framework for VI-NES behavior. This improves design strategies for enhanced energy dissipation and robustness in real-world applications.
\end{abstract}

\begin{keywords}
nonlinear energy sink, friction, slow invariant manifold, multiple scales method, nonlinear mode, impact map, bifurcation, strongly modulated response, frequency response curve. 
\end{keywords}

\section{Introduction} \label{section: Introduction}
This paper presents a unified analytical and numerical framework for characterizing the dynamics of a symmetric vibro-impact nonlinear energy sink with dry friction. By extending the multiple scales method to include friction and combining it with an impact-map approach, closed-form expressions for performance metrics are derived linking tuning parameters, including friction, to activation, dissipation, and stability within the two-impacts-per-period regime. The analysis reveals coexisting impact- and sliding-dominated responses near the optimal operating range and clarifies the amplitude-dependent role of friction in targeted energy transfer.
The design of mechanical structures subjected to dynamic loads, such as turbines, aircraft, civil engineering structures, and rotating machinery, relies heavily on understanding and controlling vibrations. External forcing or dynamical interactions between components can often induce undesired vibratory behavior potentially leading to malfunctions or total failure of mechanical structures~\cite{nucera2007targeted,lu2018particle}. Given these risks, structural vibration control has become a focal point in ensuring the reliability and longevity of mechanical systems~\cite{gourc2015quenching,rahman2015performance}. Whether through modifying the physical properties of the system, e.g. added damping, or through active  or passive control~\cite{eissa2006comparison,pratt1999terfenol,sayed20121,el2015passive,vakakis2018passive}, all strategies and techniques aim to dissipate unwanted vibrational energy efficiently.
A widely studied passive energy control strategy is targeted energy transfer (TET)~\cite{vakakis2022nonlinear,vakakis2018passive}, which directs vibrational energy irreversibly from a primary structure to a nonlinear light attachment known as a nonlinear energy sink (NES).
NES configurations have been continuously optimized to improve energy-transfer efficiency and broaden their effective operating range. Among these, bistable NES (BNES) designs using negative-stiffness elements have demonstrated high efficiency in vibration absorption~\cite{al2014highly,romeo2015dynamics}. More recent studies combining bistability with vibro-impact mechanisms have shown enhanced performance, improved robustness, and extended operational bandwidth~\cite{LI2022103891,wang2022vertical}.
The use of BNES, however, is restricted to devices in which negative stiffness effects can be realized.\\
A particularly effective class of NES, especially for high-amplitude oscillations where traditional damping methods fall short, is the vibro-impact nonlinear energy sink (VI-NES)~\cite{al2021comparison,li2022dynamics}. This absorber mitigates vibrations primarily through impacts, attenuating both stationary and transient responses over a broad frequency range.\\
A typical VI-NES consists of a lightweight auxiliary mass that moves freely within a cavity and collides with its boundaries. This concept has been explored in various configurations~\cite{al2021comparison,farid2021dynamics}, including single- and double-sided designs~\cite{al2013numerical,lizunov2024comparison,lin2025dynamic}, inclined~\cite{zhang2015periodic,zhang2019stability,heiman1987dynamics}, and rotational variants~\cite{saeed2020rotary,hwang2007rotational}, as well as setups incorporating additional cubic nonlinearities~\cite{sayed20121,wu2023targeted}. Due to their consistent performance across diverse excitation scenarios~\cite{chatterjee1996impact,gendelman2011targeted}, VI-NES have also been adapted to more complex setups and practical applications, such as seismic mitigation~\cite{nucera2007targeted,al2013numerical}, vibration suppression under traveling loads~\cite{li2025effectiveness}, and vibration-based energy harvesting~\cite{li2022dynamics,li2024electromagnetic}.
Most prior studies have primarily focused on identifying the most effective operating conditions for TET \cite{gendelman2001energy,vakakis2001energy}, by analyzing a frictionless, symmetric VI-NES exhibiting two symmetric impacts per excitation cycle near primary resonance. This idealized configuration serves as a logical starting point, as the absence of friction and the symmetry properties of the VI-NES simplify the analysis by isolating the role of impacts on system behavior and allowing a clear analytical characterization of the most effective response regime that marks the onset of TET and the activation of the VI-NES. Optimal design rules have been derived for these cases \cite{YOUSSEF2021116043,qiu2019design,lizunov2023selection,li2024design}, providing guidelines for selecting key VI-NES parameters, such as mass ratio, cavity length, and coefficient of restitution, based on the oscillation amplitude and frequency range. 
However, such idealized models fail to capture the complexities of real-world behavior, where friction plays a crucial role in shaping the VI-NES response. Previous studies that included friction have only examined specific configurations or investigated the frictional effects to a limited extent, either experimentally or through numerical parametric analyses \cite{theurich2019effects,heiman1988periodic,zhang2015periodic,fan2018analysis,li2021importance,liu2018stochastic}. Consequently, a systematic analytical framework that quantifies the influence of friction on response regimes, its interaction with other tuning parameters, and its impact on overall VI-NES performance is still lacking.
In practice, the presence of friction introduces additional response regimes, which are critical for system design and optimization. Specifically, the interaction between impact-dominated and sliding-dominated behaviors remains insufficiently explored, despite its significant influence on energy-transfer efficiency and absorber robustness.\\
This study aims to enhance the predictability of energy dissipation in vibro-impact systems by analyzing the dynamics of a linear oscillator (LO) coupled to a symmetric, horizontally oriented VI-NES subjected to dry friction. The coupling is governed by both impact and friction forces, and the work focuses on their combined influence on system behavior. Building upon the analytical framework developed in our previous study of the symmetric frictionless VI-NES~\cite{YOUSSEF2021116043}, the present work extends the Multiple Scales Method (MSM)~\cite{YOUSSEF2021116043,lakrad2002periodic,el2015passive,li2022dynamics,maaita2013effect} to explicitly account for frictional effects. This extension enables the derivation of closed-form expressions for design-relevant performance metrics, thereby providing a new analytical approach for evaluating VI-NES performance under realistic conditions. To complement the MSM, an impact-map approach~\cite{leine2012global,han1995chaotic,heiman1988periodic} is employed to capture the fast dynamics, determine the stability boundaries of the two-impacts-per-period ($2$-IPP) regime, and fully characterize the optimal operating range of the frictional VI-NES. The global behavior, including the emergence of higher-order impact sequences and the associated bifurcations, was investigated in a recent study~\cite{youssef2025asymmetric} for the general case of an asymmetric VI-NES with dry friction. Together, these contributions establish a comprehensive analytical and numerical framework that deepens the understanding of VI-NES dynamics, clarifies the interplay between friction, impacts, and tuning parameters, and provides practical tools for the design and optimization of vibro-impact nonlinear energy sinks.\\
This paper is organized as follows:  Section \ref{section: Model Description} presents the model description, detailing the modifications introduced to incorporate friction as well as the relevant dimensionless parameters. Section~\ref{section: MSM} extends MSM to frictional VI-NES, deriving the slow dynamics and SIM for symmetric $2$-IPP solutions. Section \ref{section: Impact Map} introduces the impact-map approach, emphasizing its ability to capture fast system dynamics and predict the stability and bifurcation properties of the studied periodic solutions. Section \ref{Section: response regimes} synthesizes the results from the MSM and impact-map analyses to define design-relevant metrics that quantify the performance and efficiency of the VI-NES, followed by a discussion of the coexistence between impacting and purely sliding response regimes and the approximation of the corresponding frequency response curves based on the MSM formulation. Section \ref{section: Conlusion} concludes with key findings and future research directions for optimizing VI-NES performance in practical applications.
\section{Model description}\label{section: Model Description}
\begin{figure}[b!]
	\centering
	\begin{subfigure}[b]{0.475\textwidth}
		\centering
		\includegraphics[width=\linewidth]{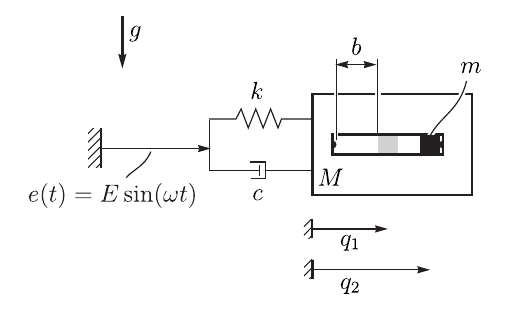}
		\caption{Model scheme of a LO VI-NES.}\label{fig:model LOVINES}
	\end{subfigure} \quad
	\begin{subfigure}[b]{0.475\textwidth}
		\centering
		\includegraphics[width=\linewidth]{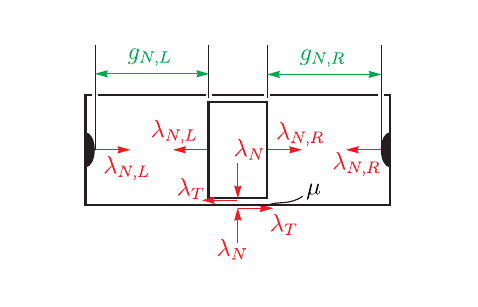}
		\caption{Free body diagram of the VI-NES.}\label{fig:model FBD LOVINES}
	\end{subfigure}
	\caption{Linear oscillator (LO) excited by the base motion of $e(t)$ and undergoing dry friction and impacts from its coupling with a vibro-impact absorber (VI-NES). }
	\label{fig:model}
\end{figure}

The mechanical system under consideration consists of a single degree of freedom damped linear oscillator coupled with a symmetric VI-NES and is illustrated in Figure (\ref{fig:model LOVINES}). The main structure, represented by the linear oscillator of mass $M$, damping coefficient $c$ and stiffness $k$, is subjected to a harmonic base excitation of amplitude $E$ and frequency $\omega$.  The absolute displacements of the primary structure and the auxiliary mass, are represented by the coordinates $q_1$ and $q_2$, respectively. 
The system is modeled within a rigid-body framework, where the auxiliary mass~$m$, constrained to horizontal motion through bilateral frictional constraints, is subjected to dry friction at both the top and bottom surfaces of the cavity. Since these frictional forces act in the same direction to resist the relative motion of the auxiliary mass, they can be combined into a single effective force, as shown in the free body diagram in Figure (\ref{fig:model FBD LOVINES}). During the movement of the primary structure, the auxiliary mass $m$ may also impact the internal side walls of the cavity.
The contact between the absorber and the side walls of the cavity is modeled by Signorini's law, which ensures that the auxiliary mass does not penetrate the side walls of the cavity, when in contact. Herein, the force law between the signed contact gap functions $g_{N,L}=q_1-q_2+b$ for the left wall and $g_{N,R}=-\left(q_1-q_2\right)+b $ for the right wall and the compressive normal forces $\lambda_{N,L}$ and $\lambda_{N,R}$ is given by Signorini's law \cite{leine2013dynamics}
\begin{align}
\bm{0}\leq \bm{\lambda}_{N} \perp \bm{g}_N \geq \bm{0} \, , \quad
\bm{\lambda}_{N} =\left(\lambda_{N,L} \quad \lambda_{N,R}\right)^\textrm{T}  \, ,
\quad \bm{g}_N=\left(g_{N,L} \quad g_{N,R}\right)^\textrm{T} \, .	 \label{eq: Signorini's law}
\end{align} 
In this context, the parameter $b$ represents the distance between the middle of the cavity and either sidewalls and is therefore used as a nominal characteristic distance in the upcoming derivations. The inequality complementarity condition imposed by (\ref{eq: Signorini's law}) can be explained as follows: As soon as the absorber comes in contact with the right wall of the cavity, i.e., the relative displacement of the absorber in the cavity is $q_1-q_2=b$ (accordingly $g_{N,R}=0$), a normal force $\lambda_{N,R}\geq0$ acts between the mass $m$ and the wall to avoid penetration. If the contact at the right wall is however open, i.e., $q_1-q_2<b$ (accordingly $g_{N,R}>0$), then $\lambda_{N,R}=0$. The same principle applies to the left wall of the cavity, where contact occurs for $q_1-q_2=-b$, resulting in a normal force $\lambda_{N,L}\geq0$ to prevent penetration, and an open contact, i.e., $q_1-q_2>-b$, means that $\lambda_{N,L}=0$.\\
The contact between the auxiliary mass $m$ and the lower surface of the cavity is assumed to be always closed with a constant non-zero normal force $\lambda_{N}=mg$, which is due to the constant gravitational field always acting on the absorber. In this work, the Coulomb law is adopted to model the dry friction between the absorber and the cavity surface, where the friction coefficient~$\mu$ is assumed to be identical for both directions of sliding. The frictional force is denoted by $\lambda_T$ and the corresponding law written in primal and dual form as a normal cone inclusion reads as \cite{leine2013dynamics}
\begin{align}
-\lambda_T \in \mu \lambda_{N}\Sign \left(\gamma_T\right) \quad \Leftrightarrow \quad \gamma_T \in \mathcal{N}_{C_T}(-\lambda_T) \, , \label{eq: Coulomb's law}
\end{align}
where $\gamma_T$ denotes the relative contact velocity and is defined as $\gamma_T=\dot{q}_1-\dot{q}_2$ and the negative force reservoir $C_T$ is given by the interval $\left[-\mu m g \, , \, +\mu m g \right]$.\\
The occurring impacts on sidewalls of the cavity are characterized by the same Newtonian coefficient of restitution $r \in \left[ 0 , 1\right]$, with the impulsive dynamics governed by the generalized Newtonian impact law
\begin{align}
0 \leq \gamma_{N,i}^+ +r \gamma_{N,i}^-\perp \Lambda_{N,i} \geq 0 \, \,  , \quad  i=R,L  \, ,\label{eq: Newton's impact laws}
\end{align}
where $\Lambda_{N,i}$ denotes the impulsive force and the post- and pre-impact contact velocities at the collision time instant $t^{c}$ are given by 
\begin{equation}
\gamma_{N,i}^+=\lim_{t \downarrow t^{c}}\dot{g}_{N,i}(t)\,, \quad \gamma_{N,i}^-=\lim_{t \uparrow t^{c}}\dot{g}_{N,i}(t) \, .
\end{equation}
The superscripts $(\cdot)^-$ and $(\cdot)^+$, denote the values at the time instant right before and right after the impact, respectively, and a dot above a variable represents the differentiation with respect to time $t$. The complementarity requirement in (\ref{eq: Newton's impact laws}) guarantees a velocity jump at the moment of impact with a positive impulse preventing the adhesion of the absorber to the side walls of the cavity. Choosing a coefficient of restitution $0 \leq r<1$, enhances the dissipative character of the VI-NES and adds to the frictional dissipation between impacts a dissipation of energy with every occurring impact.\\
In the following derivations, phases of persistent contact of the absorber with the sidewalls, or its sticking to the surface of the cavity, will not be considered as they are irrelevant for the response regime of interest. The non-impulsive dynamics of the system,  between two consecutive impacts, can therefore be described by 
\begin{align}
\begin{split}
|q_1-q_2|<b : \quad &	\quad M \ddot{q}_1+c \dot{q}_1+k q_1 = k e(t) + c \dot{e}(t) - \mu m g \, \sign (\dot{q}_1-\dot{q}_2)  \quad \text{a.e.}\\
&	 \quad  m \ddot{q}_2= \mu m g \,\sign (\dot{q}_1-\dot{q}_2)  \quad \text{a.e.}
\end{split}\label{eq: eq of motion 1a}
\end{align}
where $\dot{q}_1\neq\dot{q}_2$. The time instants for which $\dot{q}_1=\dot{q}_2$ are assumed to be Lebesque negligible and therefore (\ref{eq: eq of motion 1a}) holds almost everywhere (a.e.).
The impulsive dynamics governed by the Newtonian impact law and the conservation of linear momentum is given by 
\begin{align}
\begin{split}
|q_1-q_2|=b :\quad & \quad \left(\dot{q}_1^+ -\dot{q}_2^+\right)=-r \left(\dot{q}_1^- -\dot{q}_2^-\right)\\
& \quad   M \dot{q}_1^+ + m\dot{q}_2^+=  M \dot{q}_1^- + m\dot{q}_2^-  \, .
\end{split} \label{eq: eq of motion 1b} 
\end{align}
Equations (\ref{eq: eq of motion 1a})-(\ref{eq: eq of motion 1b}) are transformed into a normalized form to describe the dynamics of the barycentric coordinates $v$ and $w$, denoting the displacement of the center of mass and the relative displacement of the absorber within the cavity, respectively. The same steps as in the previous work \cite{YOUSSEF2021116043} are followed. Starting with the normalization of the time and coordinates according to 
\begin{equation}
\tau=\omega_0 t \, ,\quad	{\tilde{q}}_1=\frac{q_1}{b} \, ,\quad {\tilde{q}}_2=\frac{q_2}{b} \, ,\quad
v={\tilde{q}}_1+\epsilon {\tilde{q}}_2 \, ,\quad w={\tilde{q}}_1-{\tilde{q}}_2 \, , \label{eq: normalization}
\end{equation} 
where the normalized system parameters for a harmonic base excitation $e(t)=E\sin(\omega t)$ are given by
\begin{equation}
\epsilon=\frac{m}{M} ,\; \; \tilde{E}=\frac{E}{b} ,\; \; G=\frac{\tilde{E}}{\epsilon},\: \; \omega_0=\sqrt{\frac{k}{M}}, \; \; \Omega=\frac{\omega}{\omega_0} ,\; \; \lambda=\frac{c}{m \omega_0} , \; \; \tilde{g}=\frac{g}{b \omega_0^2}\, ,\label{eq: parameter normalisation}	
\end{equation}
the following transformed equations of motion can be obtained
\begin{align}
\begin{split}
&	|w|<1 :  \\
&	v^{\prime \prime}+\epsilon w^{\prime \prime}+\epsilon \lambda v^{\prime}+ v+\epsilon w+\mathcal{O}\left(\epsilon^2\right) =\epsilon G \sin(\Omega \tau) -\epsilon \mu \tilde{g}\, \sign(w^{\prime})+\mathcal{O}\left(\epsilon^2\right) \; \text{a.e.}\\
& v^{\prime \prime}  -	w^{\prime \prime}=(1+\epsilon) \mu \tilde{g}\, \sign(w^{\prime}) \quad \text{a.e.}
\end{split} \label{eq: eq of motion 2a}	\\
\begin{split}
&|w|=1 :\\
&  v^+=v^- \, , \quad  	v^{\prime+} =v^{\prime-}   \\
& w^+=w^- \, ,	 \quad w^{\prime+} =-r w^{\prime-} \, ,
\end{split} \label{eq: eq of motion 2b}
\end{align}
where the prime symbol $\left( \cdot\right)^\prime$ represents the differentiation w.r.t. the dimensionless time $\tau$.\\
It has been shown in \cite{YOUSSEF2021116043} that the various VI-NES parameters affect the performance of the damper, its operating range, as well as the stability of certain response regimes. This work investigates mainly the effect of friction on the dynamics of the studied system in the vicinity of the primary resonance ($\omega\approx \omega_0 $) aiming to determine, for a given mass ratio ($\epsilon \approx 6 $\textperthousand), the parameter set for an optimal performance.

\section{Multiple scales method} \label{section: MSM}
This section extends the results of \cite{YOUSSEF2021116043} to include dry friction in the application of the MSM to vibro-impact nonlinear energy sinks. The first-order approximation of the periodic solutions of system (\ref{eq: eq of motion 2a})-(\ref{eq: eq of motion 2b}) is obtained using the multiple scales method. The final expressions follow the same derivation steps as in \cite{YOUSSEF2021116043} and are omitted here for brevity.
The periodic solutions are expressed as expansions in terms of the mass ratio $\epsilon$ as follows
\begin{align}
v(\tau, \epsilon)&=v(\tau_0, \tau_1, \epsilon) \sim v_0(\tau_0, \tau_1)+\epsilon v_1(\tau_0, \tau_1)+\mathcal{O} (\epsilon^2)\, ,\label{v msm}\\
w(\tau, \epsilon)&=w(\tau_0, \tau_1, \epsilon) \sim w_0(\tau_0, \tau_1)+\epsilon w_1(\tau_0, \tau_1)+\mathcal{O} (\epsilon^2) \, ,\label{w msm}
\end{align}
where the fast and slow time scales are introduced as~$\tau_0=\tau$ and $\tau_1=\epsilon \tau$, respectively. Adopting the same notations as in \cite{YOUSSEF2021116043}, where $D_0$ and $D_1$ represent the partial derivatives w.r.t time $\tau_0$ and $\tau_1$, respectively, the term in the sign function in (\ref{eq: eq of motion 2a}) is expressed as
\begin{align}
\sign\left(w^\prime\right)= \sign \left(D_0 w_0 +\epsilon \left(D_0 w_1+D_1 w_0\right)+\mathcal{O}(\epsilon^2)\right)\, . 
\end{align} 
Since the motion of interest excludes the sticking of the auxiliary mass to the cavity surface, phases where $w^\prime=0$ are omitted. This is valid under the assumption $|D_0 w_0|>\mathcal{O}(\epsilon)$, which allows the relation 
\begin{equation}
\sign(w^\prime)=\sign(D_0w_0)\, , \quad \text{for} \quad |D_0 w_0|>\mathcal{O}(\epsilon)\label{eq: MSM assumption}
\end{equation}
to be established. This assumption will be consistently considered throughout the multiple scales analysis. It allows the set of ODEs (\ref{eq: eq of motion 2a})-(\ref{eq: eq of motion 2b}), which are transformed into a PDE by the multiple scales ansatz (\ref{v msm})-(\ref{w msm}), to be separated for different $\epsilon$-orders in a cascaded set of ODEs\\
\textbf{Equating terms of order $\epsilon^0$:}
\begin{align}
\begin{split}
& |w_0|<1 \, , \, \,  |D_0 w_0|>\mathcal{O}(\epsilon) : \\
& D_0^2 v_0+v_0=0 \, ,\\
& D_0^2 w_0-D_0^2v_0=-\mu \tilde{g} \sign(D_0w_0) \, ,
\end{split}
\label{eq: v0 w0 DGL} \\
\begin{split}
& |w_0|=1 : \\
& v_0^+=v_0^- \,, \quad  D_0^+ v_0 =D_0^- v_0\, , \\
& w_0^+=w_0^- 	 \, , \quad  D_0^+ w_0 =-r D_0^- w_0 \, .
\end{split} \label{eq: v0 w0 impactconditions}
\end{align}
\textbf{Equating terms of order $\epsilon^1$:}
\begin{align}
\begin{split}
&|w_0+\epsilon w_1|<1 : 	\\
&  D_0^2 v_1+v_1=-2D_0D_1v_0-\lambda D_0v_0+v_0-w_0+G \sin(\Omega \tau_0 )\, , \\
& D_0^2 w_1+v_1=-2D_0D_1w_0-\lambda D_0v_0+v_0-w_0+G \sin(\Omega \tau_0)-\mu \tilde{g} \sign(D_0 w_0)  \, .
\end{split} \label{eq: v1 w1 DGL} \\
\begin{split}
&|w_0+\epsilon w_1|=1 :\\
& v_1^+=v_1^- \, , \quad D_0^+ v_1 + D_1^+ v_0 =D_0^- v_1 + D_1^- v_0	\, ,  \\
& w_1^+=w_1^- \, , \quad  D_0^+ w_1 + D_1^+ w_0 =-r(D_0^- w_1 + D_1^- w_0) \, .
\end{split} \label{eq: v1 w1 impactconditions}
\end{align}
The general solutions of (\ref{eq: v0 w0 DGL})-(\ref{eq: v0 w0 impactconditions}) can be written in the form 
\begin{align}
v_0(\tau_0, \tau_1)&=C(\tau_1) \sin (\tau_0+\theta(\tau_1)) \label{eq: v0} \, , \\
w_0(\tau_0, \tau_1)&= C(\tau_1) \sin (\tau_0+\theta(\tau_1))+\Phi\left(\tau_0, \tau_1\right) \label{eq: w0} \, ,
\end{align}	
where the function $\Phi\left(\tau_0, \tau_1\right)$ is a continuous piecewise quadratic function w.r.t.\ the fast time scale~$\tau_0$ and its parameters are expressed as functions of the slower time scale $\tau_1$. Compared to \cite{YOUSSEF2021116043}, the equations (\ref{eq: v0 w0 DGL})-(\ref{eq: w0}) are identical but now also include additional terms in~(\ref{eq: v0 w0 DGL}) and~(\ref{eq: v1 w1 DGL}) proportional to $\mu$ which account for the friction between the auxiliary and primary mass.

\subsection{Determination of periodic motion}
\begin{figure}[ht!]
	\centering
	\begin{minipage}{0.475\textwidth}
		\centering
		\includegraphics[width=\linewidth]{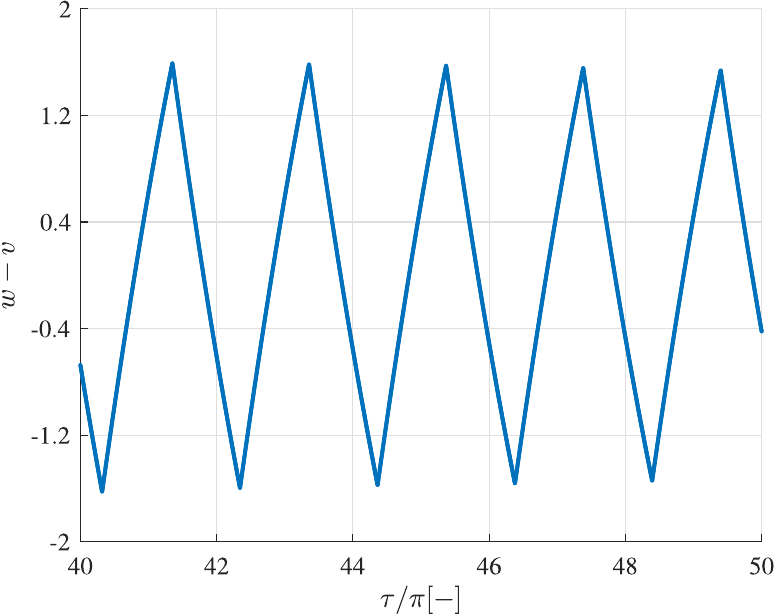}
		\caption{An example of the sought motion during a free response of system (\ref{eq: eq of motion 2a})-(\ref{eq: eq of motion 2b}) \linebreak for $\epsilon=0.006$, $\tilde{g}=0.211$, $r=0.76$ and $\mu=0.5$.}
		\label{fig:symmetric response example}
	\end{minipage}
	\hfill
	\begin{minipage}{0.475\textwidth}
		\centering
		\includegraphics[width=\linewidth]{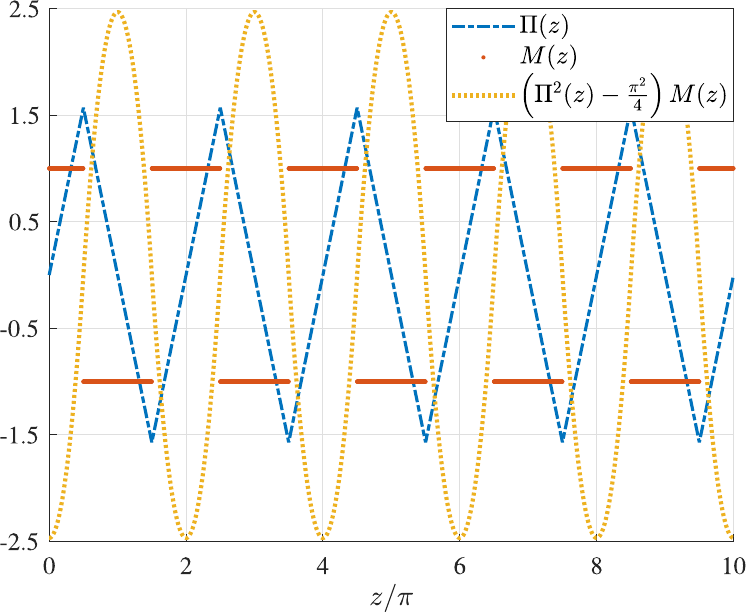}
		\caption{Representation of the nonsmooth sawtooth function $\Pi(z)$, its derivative $M(z)$ and the function $\left(\Pi^2(z)-\frac{\pi^2}{4}\right)M(z)$.}
		\label{fig:Pi_function}
	\end{minipage}
\end{figure}

The determination of the unknown parameters and functions in the solutions (\ref{eq: v0}) and (\ref{eq: w0}) relies on the studied periodic motion. In this work, the 1:1 internal resonance with two symmetric impacts per period is considered. The symmetric pattern in which the absorber hits the side walls of the cavity twice per period, imposes a sign change on the relative velocity after each impact. This can be mathematically expressed using the nonsmooth sawtooth function $\Pi(z)$ and its derivative $M(z)$ from \cite{YOUSSEF2021116043}, given by 
\begin{align}
\Pi(z)&=\arcsin(\sin(z))=\frac{4}{\pi}\sum_{k=1}^{\infty} \frac{(-1)^{k+1}}{(2k-1)^2} \sin((2k-1)z) \, , \label{eq: pi function}\\
M(z)&=\frac{\text{d} \Pi(z)}{\text{d}z}= \text{sign}(\cos(z)) \, ,\quad \forall \, z\neq\frac{\pi}{2}+k\pi \, , \quad k \in \mathbb{Z}\, . \label{eq: M function}
\end{align}
The second equation from (\ref{eq: v0 w0 DGL}) yields 
\begin{align}
D_0^2 w_0(\tau_0, \tau_1)-D_0^2v_0(\tau_0, \tau_1)&=-\mu \tilde{g} M \left(\tau_0+\eta(\tau_1)\right) \, ,\\	
\Rightarrow D_0 w_0(\tau_0, \tau_1)-D_0v_0(\tau_0, \tau_1)&=-\mu \tilde{g} \Pi \left(\tau_0+\eta(\tau_1)\right)+B(\tau_1) M\left(\tau_0+\eta(\tau_1)\right) \, ,
\end{align}
where $\tau^c_{0,k}=\frac{\pi}{2}+k\pi-\eta$ defines the collision time instants and the slow parameter $B(\tau_1)>0$ accounts for velocity jumps.
Considering the continuity of the solution $w_0$ as per (\ref{eq: v0 w0 impactconditions}), the function $\Phi (\tau_0, \, \tau_1)$ from (\ref{eq: w0}) takes the form
\begin{align}
\Phi(\tau_0, \tau_1)=B(\tau_1) \Pi\left(\tau_0+\eta(\tau_1)\right) -\frac{1}{2} \mu \tilde{g} \left(\Pi^2\left(\tau_0+\eta(\tau_1)\right)-\frac{\pi^2}{4}\right) M\left(\tau_0+\eta(\tau_1)\right). \label{eq: w ansatz}
\end{align}
An example of the function $\Phi$ is shown in Figure (\ref{fig:symmetric response example}), while its component functions are depicted in Figure (\ref{fig:Pi_function}).
The ansatz for the solution $w_0$, given by (\ref{eq: w0}) and (\ref{eq: w ansatz}), is only valid under the assumption (\ref{eq: MSM assumption}), which can be rewritten in terms of the slow variables as
\begin{align}
0<	C<B-\frac{\pi}{2}\mu \tilde{g}+\mathcal{O}\left(\epsilon\right) \, . \label{eq: msm validity range}
\end{align}
This inequality defines the range within which the multiple scales approximation is valid.
Furthermore, the dynamics will be studied for a normalized frequency close to unity, i.e.,  $\Omega=1+\sigma \epsilon$, where the parameter $\sigma$ quantifies the nearness to resonance. This assumption is relevant for the identification and removal of secular terms from the next order of approximation. The vanishing of the secular terms in the first equation of (\ref{eq: v1 w1 DGL}) yields the following set of ODEs describing the modulation of the amplitudes and phases between two consecutive impacts during the motion of interest
\begin{align}
D_1 C&= -\frac{1}{2} \left(\lambda C +\frac{4}{\pi}B \sin \left(\eta-\theta\right) - \frac{16-\pi^4}{2\pi^3} \mu \tilde{g} \cos\left(\eta - \theta \right) - G \sin \left(\sigma\tau_1-\theta\right)\right) \, ,\label{eq: D1C} \\
D_1 \theta &= -\frac{1}{2 C} \left(- \frac{4}{\pi}B \cos \left(\eta-\theta\right)-\frac{16-\pi^4}{2\pi^3} \mu \tilde{g} \sin\left(\eta - \theta \right) + G \cos \left(\sigma\tau_1-\theta\right)\right) \, .\label{eq: D1 theta}
\end{align} 
Note that the used method transforms the problem by changing the state variables to slowly changing state variables: the state of the system is described by its slowly changing amplitude and phase instead of its displacement and velocity. Moreover, it allows the study of both transient and steady-state dynamics, making it a suitable tool for understanding and analyzing the dynamics of the system at hand.
\subsection{The slow invariant manifold}
\begin{figure}[b!]
	\centering
	\begin{minipage}{0.575\textwidth}
		\includegraphics[width=\linewidth]{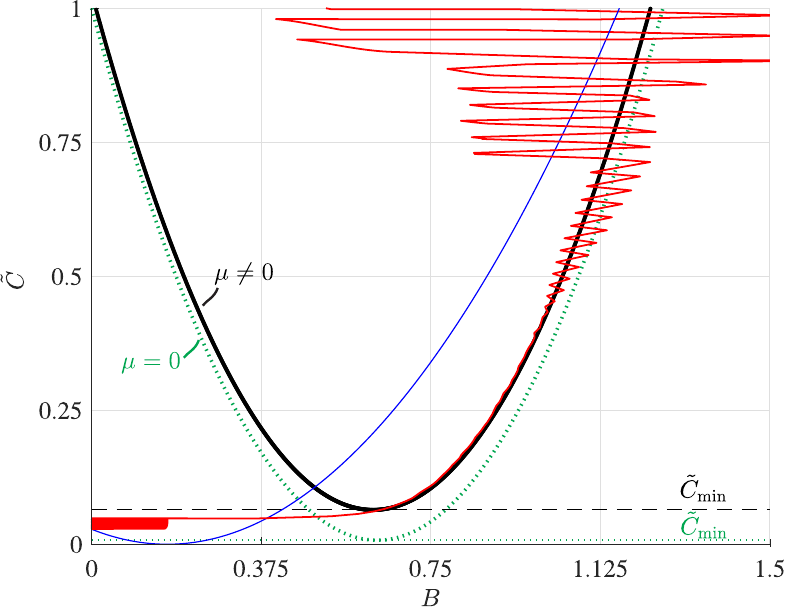}	
	\end{minipage}
	\caption{Projection of the free resonant motion of the LO VI-NES simulated using (\ref{eq: eq of motion 2a})-(\ref{eq: eq of motion 2b}) (red line) on the SIM given by (\ref{eq: SIM}) (black line) for $r =0.76$, $\mu=0.3$ and $\tilde{g}=0.355$. The dotted green line represents the SIM for $\mu=0$ and the region below the blue line corresponds to the validity region of the MSM ansatz from (\ref{eq: w ansatz}).}
	\label{fig:SIM}	
\end{figure}

To further characterize the type of motion under study, i.e., two symmetric impacts per period, the contact condition and impact equations from (\ref{eq: v0 w0 impactconditions}) are reformulated in terms of the slow parameters $\left(C,\, \theta, \, B, \, \eta\right)$ at the collision time instant $\tau^c_{0,k}=\frac{\pi}{2}+k\pi-\eta$. The first three impact equations from~(\ref{eq: v0 w0 impactconditions}) define the continuous character of the amplitude $C$, the phases $\theta$ and $\eta$ as well as the average absolute velocity $B$, while the last equation along with the contact condition yield
\begin{align}
&\textbf{Contact condition:} \;	\left| w_0(\tau^c_{0,k})\right|=1 &\Rightarrow C \cos\left(\eta-\theta\right)&=1-\frac{\pi}{2} B \, , \label{eq: SIM c1}\\	
&\textbf{Impact equation:} \; D_0^+ w_0 =-r D_0^- w_0 &\Rightarrow C \sin\left(\eta-\theta\right)&= \frac{1-r}{1+r} B +\frac{\pi}{2} \mu \tilde{g} \, . \label{eq: SIM c2}
\end{align}
Combining the above equations (\ref{eq: SIM c1}) and (\ref{eq: SIM c2}) with the trigonometric identity yields the expression for the slow invariant manifold (SIM)
\begin{align}
\tilde{C}=\left(1-\frac{\pi}{2}B\right)^2+\left(R B +h\right)^2 \quad \text{with} \; \tilde{C}=C^2 \, , \;  R=\frac{1-r}{1+r} \; \text{and} \; h=\frac{\pi}{2} \mu \tilde{g} \, . \label{eq: SIM}
\end{align}
An illustration of the SIM is presented in Figure (\ref{fig:SIM}). Similar to the frictionless analysis in \cite{YOUSSEF2021116043}, the topological structure of the SIM is maintained in the presence of dry friction. The SIM is composed of two branches: an unstable left branch and a partially stable right branch. However, the activation threshold of the VI-NES, allowing an active TET and represented by the point $\left(B_{\min}, \, \tilde{C}_{\min}\right)$ in the slow parameter space, depends not only on the coefficient $R$, but also on~$h$ (and therefore the friction coefficient $\mu$, the gravity $g$, and the cavity length $b$) and is given by
\begin{equation}
B_{\min}\left(R,h\right)= \frac{2\pi-4Rh}{\pi^2+4R^2}\, , \quad \tilde{C}_{\min}\left(R,h\right)=\frac{\left(\pi h+2R\right)^2}{\pi^2+4R^2} \, , \label{eq: B_min C_tilde_min MSM}
\end{equation}
in which it is assumed that $h<\frac{\pi}{2R}$. A projection of a simulated free resonant motion of the LO VI-NES on the SIM, shows that the stable branch of the SIM lies within the validity range for the MSM ansatz (\ref{eq: msm validity range}). 
A thorough analysis of the stability of the different solution branches follows in Section \ref{section: Impact Map}, where the expressions of the bifurcation points on the SIM defining the lower and upper bounds of its stable branch, will be analytically established.

\subsection{Steady-state solutions and flow on the SIM}
\begin{figure}[h!]
	\centering
	\begin{subfigure}[b]{0.475\textwidth}
		\centering
		\includegraphics[width=\linewidth]{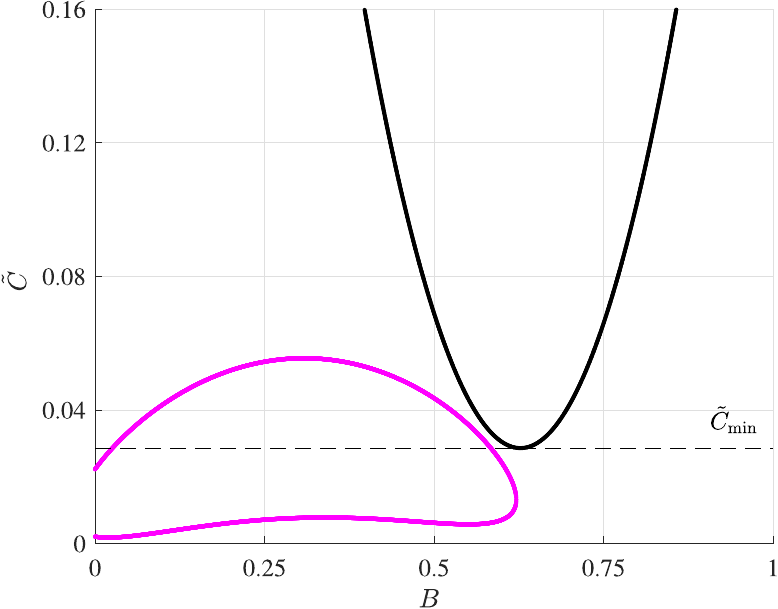}
		\caption{$\sigma=5.1$, $\Omega=1.039$}
	\end{subfigure} \quad
	\begin{subfigure}[b]{0.475\textwidth}
		\centering
		\includegraphics[width=\linewidth]{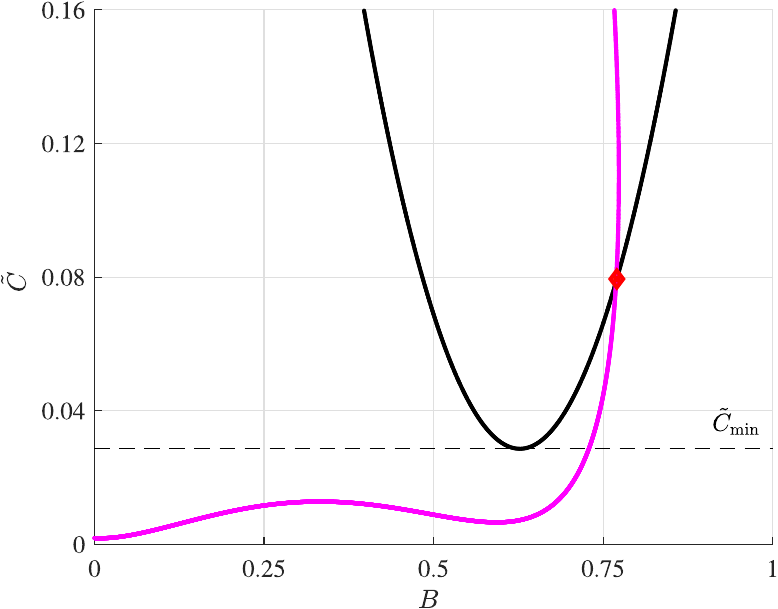}
		\caption{$\sigma=1.25$, $\Omega=1.007$}
	\end{subfigure}\\
	\begin{subfigure}[b]{0.475\textwidth}
		\centering
		\includegraphics[width=\linewidth]{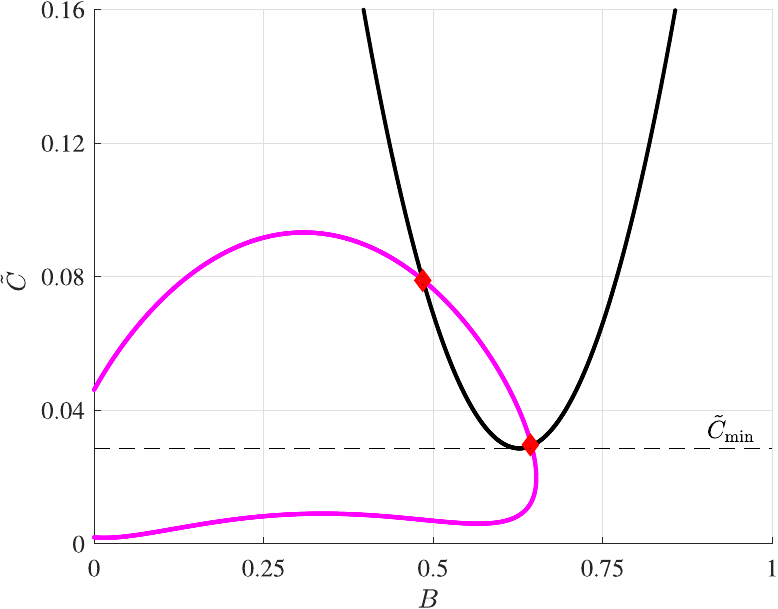}
		\caption{$\sigma=3.65$, $\Omega=1.022$}
	\end{subfigure} \quad
	\begin{subfigure}[b]{0.475\textwidth}
		\centering
		\includegraphics[width=\linewidth]{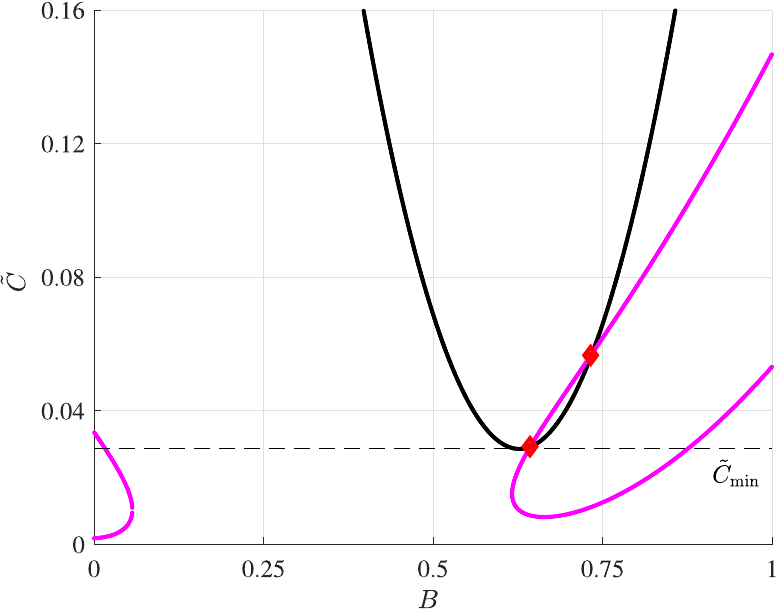}
		\caption{$\sigma=-4.4$, $\Omega=0.973$}
	\end{subfigure}
	\caption{Position of the equilibria for a level of excitation $G=1.65$ and four values of $\sigma$. The black line represents the SIM from (\ref{eq: SIM}). The purple line represent the solutions of (\ref{eq: ESIM}). The VI-NES parameters are $r =0.76$, $\mu=0.5$ and $\tilde{g}=0.105$.  }
	\label{fig:SS_schnittpunkte}	
\end{figure}

A periodic solution of (\ref{eq: eq of motion 2a})-(\ref{eq: eq of motion 2b}) with two symmetric impacts per period is characterized by its constant amplitude $C$ (or, equivalently, by its constant squared amplitude $\tilde{C}=C^2$), a constant phase difference between the response and the external excitation, hereafter denoted by $\gamma=\theta-\sigma\tau_1$ and a symmetric motion of the absorber, with a constant average absolute velocity $B$. This behavior can be described through the equilibrium of the system
\begin{align}
\begin{split}
D_1 \tilde{C} &=-\lambda \tilde{C} -\frac{\pi}{4}B \left(RB+h\right)+ \frac{16-\pi^4}{\pi^4} h \left(1-\frac{\pi}{2}B\right) - G  \sqrt{\tilde{C}}\sin \left(\gamma\right)   \\
D_1 \gamma &=\frac{1}{2\tilde{C}} \left(\frac{4}{\pi} B \left(1-\frac{\pi}{2}B\right)+ \frac{16-\pi^4}{\pi^4}h \left(RB+h\right)-G  \sqrt{\tilde{C}}\cos \left(\gamma\right) \right)-\sigma
\end{split}\, ,
\label{eq: slow dynamics impacting absorber}
\end{align}
obtained by using (\ref{eq: SIM c1}) and (\ref{eq: SIM c2}) to replace the sine and cosine terms in (\ref{eq: D1C})-(\ref{eq: D1 theta}), fulfilling the conditions of symmetric two impacts per period ($2$-IPP) motion.
Equations (\ref{eq: slow dynamics impacting absorber}) describe the phase and amplitude dynamics and are used to determine the existence and stability properties of equilibria on the SIM, corresponding to symmetric periodic motions of the primary and auxiliary mass with $2$-IPP. Setting the right-hand side of (\ref{eq: slow dynamics impacting absorber}) to zero, as to find equilibria, reduces the problem after elimination of $\gamma$ to a single quadratic algebraic equation of the form 
\begin{align}
& \left(\lambda^2 +4\sigma^2\right) \tilde{C}^2 -\left( 2\lambda k_1(B) +4\sigma k_2(B)   +G^2\right)\tilde{C}+k_1^2(B)+k_2^2(B)=0  \label{eq: ESIM}\\
&\text{with} \quad k_1(B)=-\frac{4}{\pi}B\left(RB+h\right)+\frac{16-\pi^4}{\pi^4}h \left(1-\frac{\pi}{2}B\right) \, , \\
&\text{and} \quad k_2(B)=\frac{4}{\pi}B\left(1-\frac{\pi}{2}B\right)+\frac{16-\pi^4}{\pi^4}h \left(RB+h\right)  \, .
\end{align}
Solving (\ref{eq: ESIM}) for positive real roots delivers a closed-form analytical expression of the steady-state squared amplitude $\tilde{C}$ as a function of the velocity $B$. The obtained expression describes another manifold in the slow parameter configuration space, that unlike the SIM from (\ref{eq: SIM}), depends on the external excitation through the parameters $G$ and $\sigma$, as well as the damping constant $\lambda$. 
Thus, steady-state solutions with 1:1 resonance and 2 symmetric impacts per period can be identified graphically as the intersection of the two manifolds in the $(B, \, \tilde{C})$-plane. A graphical illustration is given in Figure (\ref{fig:SS_schnittpunkte}) for different excitation frequencies and the same level of forcing. One should note that only points on the purple curves which agree with the SIM have a physical meaning and represent the equilibria of (\ref{eq: slow dynamics impacting absorber}). The stability of the equilibria on the SIM is determined by examining the eigenvalues of the Jacobian matrix $\mathcal{J}$ of the right-hand sides of (\ref{eq: slow dynamics impacting absorber}), evaluated at the equilibrium solution, after expressing the velocity $B$ as
\begin{equation}
B(\tilde{C})=B_{\text{min}}\pm \sqrt{\frac{2}{\pi-2Rh} B_{\text{min}}}\sqrt{\tilde{C}-\tilde{C}_{\min}} \, .
\end{equation} 
The resulting slow flow on the SIM is described by 
\begin{align}
\begin{split}
D_1 \tilde{C} &=f_c(\tilde{C},\gamma)=-\tilde{\alpha}_c \tilde{C} \mp\tilde{\beta}_c\sqrt{\tilde{C}-\tilde{C}_{\min}} +\tilde{K}_c- G  \sqrt{\tilde{C}}\sin \left(\gamma\right) \\
D_1 \gamma &=f_\gamma(\tilde{C},\gamma)=\frac{1}{2\tilde{C}} \left(-\tilde{\alpha}_\gamma \tilde{C} \mp\tilde{\beta}_\gamma\sqrt{\tilde{C}-\tilde{C}_{\min}} +\tilde{K}_\gamma- G  \sqrt{\tilde{C}}\cos \left(\gamma\right) \right) 
\end{split}\, ,
\label{eq: slow dynamic impacting}
\end{align}
with the coefficients 
\begin{align}
\begin{split}
\tilde{\alpha}_c &= \lambda+\frac{4}{\pi}\frac{R}{\frac{\pi^2}{4}+R^2} \, ,\\
\tilde{\alpha}_\gamma& =\frac{2}{\frac{\pi^2}{4}+R^2}+2\sigma\, ,\\
\tilde{\beta}_c&= \frac{1}{\sqrt{\frac{\pi^2}{4}+R^2}} \left(  \frac{8}{\pi}R B_{\min} + \frac{16+8\pi^2-\pi^4}{2\pi^3} h \right)   \, , \\
\tilde{\beta}_\gamma&=  \frac{1}{\sqrt{\frac{\pi^2}{4}+R^2}}  \left(  4 B_{\min}-\frac{4}{\pi}-\frac{16-\pi^4}{\pi^4} R h  \right)     \, ,\\
\tilde{K}_c&=  -\frac{4}{\pi}R \left(B_{\min}^2-\frac{\tilde{C}_{\min}}{\frac{\pi^2}{4}+R^2}\right) - \frac{16+8\pi^2-\pi^4}{2\pi^3} h B_{\min}+\frac{16-\pi^4}{\pi^4} h\, ,\\
\tilde{K}_\gamma &=\frac{4}{\pi}B_{\min} \left(1-\frac{\pi}{2}B_{\min}\right) +\frac{2}{\frac{\pi^2}{4}+R^2} \tilde{C}_{\min} + \frac{16-\pi^4}{\pi^4}h \left(R B_{\min}+h\right)\, ,
\end{split}\label{eq: Coefficients}
\end{align}
and the corresponding Jacobian matrix 
\begin{align}
\mathcal{J}=
\begin{pmatrix}
\frac{\partial f_c}{\partial \tilde{C}}       & \frac{\partial f_c}{\partial \gamma} \\\\
\frac{\partial f_\gamma}{\partial \tilde{C}}       & \frac{\partial f_\gamma}{\partial \gamma}
\end{pmatrix} \, .
\end{align}
If both eigenvalues have a negative real part, the corresponding equilibrium is locally asymptotically stable. If an eigenvalue has a positive real part, the equilibrium is unstable. Due to the complexity of the equations above and for the sake of conciseness, the eigenvalues of the derived Jacobian matrix and the stability of the equilibria are numerically evaluated.\\
Since the SIM governs the system's long-term behavior after initial transients have decayed, the equations resulting from the MSM allow to determine the existence of periodic solutions, describe the slow flow once it reaches the SIM, and assess the stability of equilibria within it.\\
Once on the SIM, the system’s slow dynamics dictate the stability of steady-state periodic solutions, such as symmetric $2$-IPP motion. A stable equilibrium on the SIM ensures that nearby trajectories remain close, enabling sustained periodic energy transfer between the primary system and the NES. However, if the equilibrium is unstable, the slow flow diverges from it, potentially leading to chaotic or transient responses.
However, these equations do not provide any information about the attractiveness of the SIM itself. If the SIM is attractive, trajectories in its vicinity will asymptotically converge to it, even if the system starts off the SIM. 
Notably, it is possible for the slow dynamics to possess an unstable equilibrium located on the stable branch of the SIM. The behavior of the system in such cases will be discussed in Section \ref{Section: response regimes}.
The upcoming section will carry out the stability and bifurcation analysis for the SIM, identifying the set of all possible solutions of the motion under consideration and determining the corresponding attractive branches.
\section{Impact map and stability analysis} \label{section: Impact Map}
\subsection{The simplified model}
\begin{figure}[b!]
	\centering
	\begin{minipage}{0.475\textwidth}
		\includegraphics[width=\linewidth]{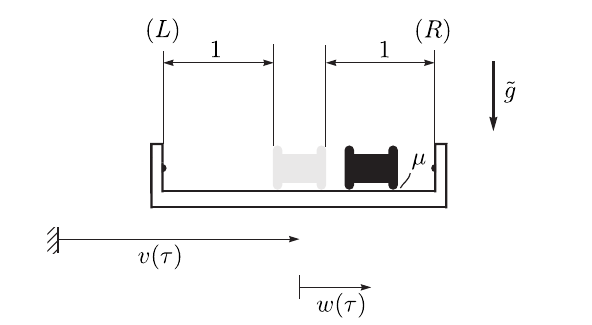}
	\end{minipage}
	\caption{Model of a harmonically excited VI-NES subjected to dry friction.}
	\label{Fig: reduced model}
\end{figure}
The following analysis focuses on cases where the primary structure oscillates harmonically with a constant amplitude at a stable steady state, driven by a harmonic input with amplitude $G$ and normalized frequency $\Omega$, that satisfies the existence condition for a stable steady-state solution of the system (\ref{eq: slow dynamics impacting absorber}). At this state, the barycentric coordinate $v$ oscillates harmonically with a constant phase $\gamma$ and amplitude $C$, unaffected by the motion of the auxiliary mass due to the small mass ratio ($\epsilon \ll 1$). Consequently, the problem reduces to a one-degree-of-freedom model, where only the motion of the absorber moving within the harmonically excited cavity is examined.
The new reduced system studied in this section is presented in Figure (\ref{Fig: reduced model}) and the corresponding equation of motion reads
\begin{align}
&|w|<1 : \nonumber\\
&	w^{\prime \prime}= v^{\prime \prime}  - (1+\epsilon) \mu \tilde{g} \, \sign(w^{\prime}) 	=-C\Omega^2 \sin\left(\Omega \tau +\gamma\right)- (1+\epsilon)\mu \tilde{g} \, \sign(w^{\prime}) \, .\label{eq: reduced eq of motion a}\\
&|w|=1 :\nonumber\\
& w^+=w^- \, ,	 \quad w^{\prime+} =-r w^{\prime-} \, .
\label{eq: reduced eq of motion b}
\end{align}
For simplicity, the values of the relative displacement and velocity of the absorber at the time instant $\tau_n$ at which the contact occurs are rewritten using the subscript $(\cdot)_n$. The reformulated contact condition and the impact law from (\ref{eq: reduced eq of motion b}) read
\begin{align}
w\left(\tau_n\right)=w_n=\left\{
\begin{array}{ll}
-1 & \text{contact on the right} \\
+1 & \text{contact on the left}
\end{array} \right.
\quad \text{and} \quad \left(w_n^\prime\right)^+=-r \left(w_n^\prime\right)^- \, .\label{eq: impact and contact cond}
\end{align}

\subsection{The elementary mappings}
Since the periodic motion of the absorber within the cavity is of main interest here, the study of the dynamics is carried out through the use of an impact map \cite{leine2012global,han1995chaotic} to depict the periodic trajectories of the impacting mass as a sequence of impact events. Each impact event is characterized by two coordinates: the collision time instant $\tau_n$ and the relative velocity $w^\prime (\tau_n)$ immediately after or immediately before the impact, denoted by $\left(w^\prime_n\right)^+$ or $\left(w_n^\prime \right)^-$, respectively.
For convenience and consistency purposes, the characterization of the $n$-th impact event is done through a new dimensionless time variable $\Psi_n=\Omega \tau_n + \gamma$ and the post-/pre-impact absolute velocity of the absorber denoted by $B_n^\pm$ and defined according to
\begin{align}
B_n^\pm=\left(v_n^\prime\right)^\pm -\left(w_n^\prime\right)^\pm =C\Omega \cos \left(\Psi_n\right)-\left(w_n^\prime\right)^\pm \label{eq: B_n^+/-} \, .
\end{align} 
Consequently, depending on the choice of characterization, the state vector may read
\begin{align}
\mathbf{x}^+_n=\left(\Psi_n, \, B_n^+\right)^\text{T} \quad \text{or} \quad \mathbf{x}^-_n=\left(\Psi_n, \, B_n^-\right)^\text{T}\, . \label{eq: state vector}
\end{align} 
In the case of two alternating impacts, two impact sequences could occur: The first impact could either be on the right side, followed by an impact on the left side (LR/RL), or it could be on the left side, followed by an impact on the right side (RL/LR). Even though both cases are symmetrical due to the symmetric properties of the VI-NES, the direction in which the mass moves within the cavity determines the sign of the friction force acting on it.
In the following, the impact sequence (LR/RL) is considered, where the post-impact state vector $\mathbf{x}^+_n=\left(\Psi_n, \, B_n^+\right)^\text{T}$ characterizes the impact that has already occurred on the left side, where the auxiliary mass initiates its motion. At $\tau_{n+1}$, the mass reaches the right side and the impact parameters are sampled in the state vector $\mathbf{x}^+_{n+1}$. After a full period of excitation, the second impact takes place at the left side again, and the corresponding state vector is denoted by $\mathbf{x}^+_{n+2}$. 
The trajectory and velocity between the first two consecutive impacts are known from the integration of (\ref{eq: reduced eq of motion a}) over the open time interval $\left(\tau_n,\, \tau_{n+1}\right)$ and substitution of (\ref{eq: impact and contact cond}) in the obtained equations yields the following implicit equations
\begin{align}
\left(w_{n+1}^\prime\right)^+&=-r\left(w_{n}^\prime\right)^+ -r C \Omega \left(  \cos \left(\Omega \tau_{n+1}+\gamma\right)-\cos\left(\Omega \tau_n +\gamma\right)\right)  \nonumber\\
&\quad -r \mu \tilde{g}\left(1+\epsilon\right) \left(\tau_{n+1}-\tau_n\right)  \label{eq: first integration}\\
w_{n+1}&= w_{n} +C \left(\sin \left(\Omega \tau_{n+1}+\gamma\right)-\sin\left(\Omega \tau_n +\gamma\right)\right) +\frac{1}{2}\mu \tilde{g}\left(1+\epsilon\right)\left(\tau_{n+1}-\tau_n\right)^2 \nonumber \\
&\quad  +\left(	\left(w_{n}^\prime\right)^+-C \Omega \cos\left(\Omega \tau_n +\gamma\right)\right)\left(\tau_{n+1}-\tau_n\right) \, ,\label{eq: second integration}
\end{align}
where $w_{n}=1$ (L) and $w_{n+1}=-1$ (R).\\
Doing the same over the open time interval $\left(\tau_{n+1},\, \tau_{n+2}\right)$ yields
\begin{align}
\left(w_{n+2}^\prime\right)^+ &=-r\left(w_{n+1}^\prime\right)^+ -r C \Omega \left(  \cos \left(\Omega \tau_{n+2}+\gamma\right)-\cos\left(\Omega \tau_{n+1} +\gamma\right)\right) \nonumber \\
& \quad +r \mu \tilde{g} \left(1+\epsilon\right) \left(\tau_{n+2}-\tau_{n+1}\right)\, , \label{eq: first integration 2} \\
w_{n+2}&=w_{n+1} +C \left(\sin \left(\Omega \tau_{n+2}+\gamma\right)-\sin\left(\Omega \tau_{n+1} +\gamma\right)\right) \nonumber \\
& \quad -\frac{1}{2}\mu \tilde{g}\left(1+\epsilon\right)\left(\tau_{n+2}-\tau_{n+1}\right)^2 \nonumber \\
&\quad +\left(	\left(w_{n+1}^\prime\right)^+-C \Omega \cos\left(\Omega \tau_{n+1} +\gamma\right)\right)\left(\tau_{n+2}-\tau_{n+1}\right) \, ,\label{eq: second integration 2}
\end{align}
where $w_{n+1}=-1$ (R) and $w_{n+2}=1$ (L).
Using (\ref{eq: B_n^+/-}) allows to express the above derived equations in terms of the post-impact coordinates.
Accordingly, the post-impact state vectors $\mathbf{x}^+_n$ and $\mathbf{x}^+_{n+1}$ characterizing the $n$-th and $(n+1)$-th impact, corresponding to the absorber moving from the left to the right side, are related through the elementary transformation $\mathbf{G}^+_{\text{LR}}$ given by Table~\ref{eq: G LR +}). Similarly, the post-impact parameters relating the $(n+1)$-th and $(n+2)$-th impact of the same cycle, from the right to the left side, are related through the elementary transformation $\mathbf{G}^+_{\text{RL}}$ as given by Table~\ref{eq: G RL +}). The derivation of the elementary transformations $\mathbf{G}^-_{\text{LR}}$ and $\mathbf{G}^-_{\text{RL}}$, representing the transformations relating the pre-impact state vectors $\mathbf{x}^-_n$, $\mathbf{x}^-_{n+1}$ and $\mathbf{x}^-_{n+2}$, follows the same approach. Specifically, equation (\ref{eq: B_n^+/-}) is expressed in terms of pre-impact variables and then, together with equation (\ref{eq: impact and contact cond}), substituted into (\ref{eq: first integration})-(\ref{eq: second integration}) and (\ref{eq: first integration 2})-(\ref{eq: second integration 2}) to derive the corresponding transformations. These transformations are given by Table~\ref{eq: G LR -}) and Table~\ref{eq: G RL -}), respectively.
\begin{table*}[t!]
	\centering
	\begin{subtable}[h]{\textwidth}  
		\centering
		\begin{align*}
		\mathbf{G}^+_{\text{LR}}\left(\mathbf{x}^+_n,\, \mathbf{x}^+_{n+1}\right)
		=\begin{pmatrix}
		B_{n+1}^+ +r B_n^+ - \left(1+r\right) C\Omega \cos\Psi_{n+1} \cdots \\\cdots -r\mu \tilde{g}\frac{1+\epsilon}{\Omega} \left(\Psi_{n+1}-\Psi_n\right) \\\\
		-\frac{B_n^+}{\Omega}\left(\Psi_{n+1}-\Psi_n\right)+C \left(\sin\Psi_{n+1}-\sin\Psi_n\right)\cdots \\ \cdots+\frac{1}{2}\mu \tilde{g}\frac{1+\epsilon}{\Omega^2}\left(\Psi_{n+1}-\Psi_n\right)^2+2 
		\end{pmatrix}=\mathbf{0}\, .
		\end{align*}
		\caption{}\label{eq: G LR +}
	\end{subtable}
	\begin{subtable}[h]{\textwidth}
		\centering
		\begin{align*}
		\mathbf{G}^+_{\text{RL}}\left(\mathbf{x}^+_{n+1},\, \mathbf{x}^+_{n+2}\right)
		=\begin{pmatrix}
		B_{n+2}^+ +r B_{n+1}^+ - \left(1+r\right) C\Omega \cos\Psi_{n+2} \cdots \\ \cdots +r\mu\tilde{g}\frac{1+\epsilon}{\Omega} \left(\Psi_{n+2}-\Psi_{n+1}\right) \\\\
		C\left(\sin\Psi_{n+2}-\sin\Psi_{n+1}\right)-\frac{B_{n+1}^+}{\Omega}\left(\Psi_{n+2}-\Psi_{n+1}\right) \cdots \\ \cdots-\frac{1}{2}\mu \tilde{g}\frac{1+\epsilon}{\Omega^2}\left(\Psi_{n+2}-\Psi_{n+1}\right)^2-2 
		\end{pmatrix} = \mathbf{0} \, .
		\end{align*}
		\caption{}\label{eq: G RL +}
	\end{subtable}
	\begin{subtable}[h]{\textwidth}
		\centering
		\begin{align*}
		\mathbf{G}^-_{\text{LR}}\left(\mathbf{x}^-_n,\, \mathbf{x}^-_{n+1}\right)
		=\begin{pmatrix}
		rB_n^-+B_{n+1}^--\left(1+r\right)C \Omega\cos\Psi_n \cdots \\ \cdots+\mu \tilde{g}\frac{1+\epsilon}{\Omega}\left(\Psi_{n+1}-\Psi_n\right)\\\\
		C \left(\sin \Psi_{n+1}-\sin \Psi_n\right) +2 \cdots \\ \cdots +\frac{1}{2}\mu \tilde{g}\frac{1+\epsilon}{\Omega^2}\left(\Psi_{n+1}-\Psi_n\right)^2		 \cdots \\ \cdots
		-\left(rB_n^- -\left(1+r\right)C \Omega \cos \Psi_n\right) \left(\frac{\Psi_{n+1}-\Psi_n}{\Omega}\right)
		\end{pmatrix}  = \mathbf{0} \, .
		\end{align*}
		\caption{}\label{eq: G LR -}
	\end{subtable}
	\begin{subtable}[h]{\textwidth}
		\centering
		\begin{align*}
		\mathbf{G}^-_{\text{RL}}\left(\mathbf{x}^-_{n+1},\, \mathbf{x}^-_{n+2}\right) 	=\begin{pmatrix}
		rB_{n+1}^-+B_{n+2}^--\left(1+r\right)C \Omega\cos\Psi_{n+1} \cdots \\ \cdots -\mu \tilde{g}\frac{1+\epsilon}{\Omega}\left(\Psi_{n+2}-\Psi_{n+1}\right) \\\\
		C \left(\sin \Psi_{n+2}-\sin \Psi_{n+1}\right) -2 \cdots\\ \cdots  -\frac{1}{2}\mu \tilde{g}\frac{1+\epsilon}{\Omega^2} \left(\Psi_{n+2}-\Psi_{n+1}\right)^2 \cdots \\ \cdots
		-\left(rB_{n+1}^- -\left(1+r\right)C \Omega \cos \Psi_{n+1}\right) \left(\frac{\Psi_{n+2}-\Psi_{n+1}}{\Omega}\right)
		\end{pmatrix}  = \mathbf{0} \, .
		\end{align*}
		\caption{}\label{eq: G RL -}
	\end{subtable}
	\caption{Elementary mappings for alternating impacts.}
	\label{tab:equations G_LR/G_RL}
\end{table*}				
The obtained elementary mappings can be iterated to generate various types of motion with alternating impacts, specifically the $2$-IPP motion examined in this work. This motion can be described by two implicit impact maps, $\mathbf{G}^+$ or $\mathbf{G}^-$,  constructed from the derived elementary mappings from Table \ref{tab:equations G_LR/G_RL} as follows:
\begin{align}
\mathbf{G}^\pm\left(\mathbf{x}^\pm_n,\, \mathbf{x}^\pm_{n+2}\right)=
\begin{pmatrix}
\mathbf{G}^\pm_{\text{LR}}\left(\mathbf{x}^\pm_n,\, \mathbf{x}^\pm_{n+1}\right) \\
\mathbf{G}^\pm_{\text{RL}}\left(\mathbf{x}^\pm_{n+1},\, \mathbf{x}^\pm_{n+2}\right) \\
\end{pmatrix} =\mathbf{0} \, . \label{eq: G +-}
\end{align}   
It is important to note that the derived elementary mappings in this subsection describe alternating impacts. However, the auxiliary mass may also impact the same sidewall twice without reaching the opposite wall. These scenarios, along with the corresponding elementary mappings, denoted by $\mathbf{G}^\pm_{\text{LL}}$ and $\mathbf{G}^\pm_{\text{RR}}$, were derived and extensively analyzed in \cite{youssef2025asymmetric} and therefore fall beyond the scope of this paper, which focuses exclusively on the~$2$-IPP motion. 

\subsection{Periodic two impacts per cycle motion: $\mathcal{P}_2^1$-orbits}
\begin{figure}[t!]
	\centering
	\begin{minipage}{0.475\textwidth}
		\centering
		\includegraphics[width=\linewidth]{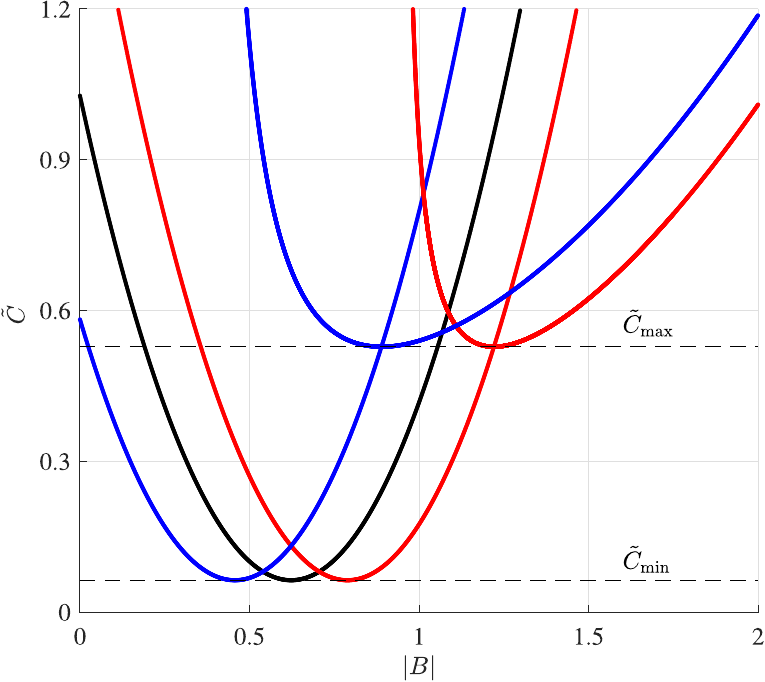}
		\caption{$\mathcal{P}_2^1$-orbits: Symmetric and asymmetric solution branches for periodic motions with 2 impacts per period for $r=0.76$, $\mu=0.1$ and $\tilde{g}=0.422$. The black, red and blue lines line are described by (\ref{eq: SIM}) and $	\mathbf{G}^{+}\left(\mathbf{x}^+_n,\, \mathbf{x}^+_{n+2}\right)=\mathbf{0}$, $ \mathbf{G}^{-}\left(\mathbf{x}^-_n,\, \mathbf{x}^-_{n+2}\right)=\mathbf{0}$, respectively. The dashed lines correspond to the excitation levels of the bifurcation points.  }
		\label{Fig: symmetric and asymmetric branches}
	\end{minipage}
	\hfill
	\begin{minipage}{0.475\textwidth}
		\centering
		\includegraphics[width=\linewidth]{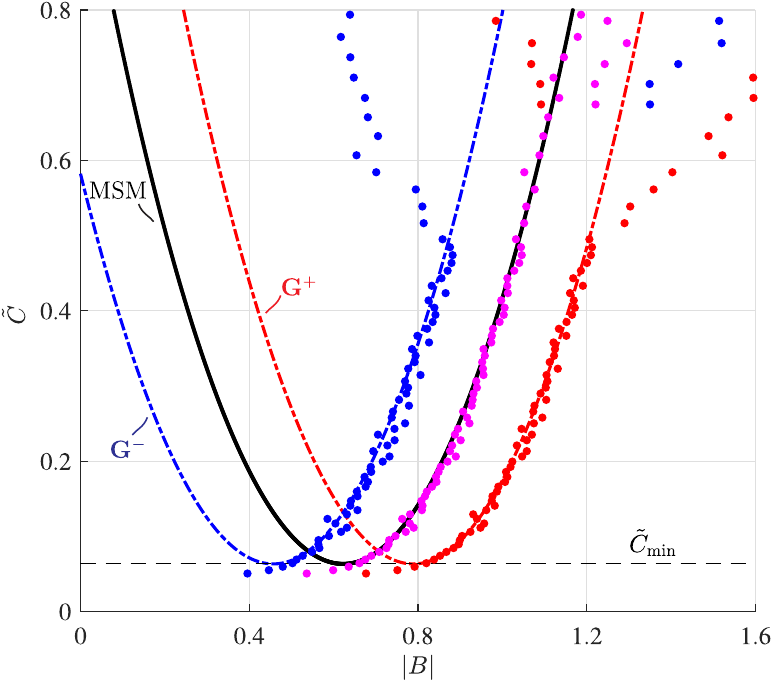}
		\caption{Projection of the free response of the LO VI-NES on the symmetric solution branches of $\mathcal{P}_2^1$-orbits for $r=0.76$, $\mu=0.1$ and $\tilde{g}=0.422$. The black, red and blue lines line are described by (\ref{eq: SIM}) and $	\mathbf{G}^{+}\left(\mathbf{x}^+_n,\, \mathbf{x}^+_{n+2}\right)=\mathbf{0}$, $ \mathbf{G}^{-}\left(\mathbf{x}^-_n,\, \mathbf{x}^-_{n+2}\right)=\mathbf{0}$, respectively. The dots correspond to the numerically computed average, post-and pre-impact velocities.}
		\label{Fig: symmetric branches and decaying flow}
	\end{minipage}
\end{figure}

The definition of an $l$-periodic orbit of order $k$, as introduced in \cite{leine2012global}, characterizes an orbit of the impact map $\mathbf{G}^\pm$ with a period of $\frac{2\pi}{\Omega} k$ and $l$ impacts occurring within one period of time ($1$:$k$ internal resonance with $l$ impacts). The $\mathcal{P}_2^1$-orbits, with two impacts per cycle in the continuous time representation, correspond to period-1 fixed points of the impact maps $\mathbf{G}^\pm$ from equation~(\ref{eq: G +-}). 
Such fixed points, in the following denoted by
\begin{equation}
\mathbf{x}_{n+i}^{\pm *}=\begin{pmatrix}	\Psi_{n+i}^*\, , \;	B_{n+i}^{\pm *} \end{pmatrix}^\text{T} \; , \qquad i=0\cdots 2\; ,
\end{equation} 
satisfy
\begin{equation}
\mathbf{G}^\pm\left(\mathbf{x}^{\pm *}_{n},\, \mathbf{x}^{\pm *}_{n+2}\right)=	\mathbf{G}^\pm\left(\mathbf{x}^{\pm *}_{n},\,
\mathbf{x}^{\pm *}_{n}\right)=\mathbf{0}\, . 
\end{equation}
This follows from the periodicity condition, which states that a 1:1 resonance condition means that the period of the system response is equal to the period of the external excitation $v(\tau)$, and that the post- or pre-impact velocities (depending on the chosen state vector from (\ref{eq: state vector})) should be equal after one response period. This condition expressed in terms of the state vectors reads
\begin{align}
\begin{pmatrix} \Psi^*_{n+2} \\ B_{n+2}^{\pm *}  \end{pmatrix}=
\begin{pmatrix} \Psi^*_n \\ B_n^{\pm *}  \end{pmatrix} +	\begin{pmatrix} 2\pi \\ 0 \end{pmatrix}  \quad \Rightarrow \quad 	\mathbf{x}_{n+2}^{\pm *}=\mathbf{x}_{n}^{\pm *}
+	\begin{pmatrix} 2\pi \\ 0 \end{pmatrix} \, .					\label{eq: periodicity condition}
\end{align} 
Inserting (\ref{eq: periodicity condition}) in (\ref{eq: G +-}) results in two sets of 4 nonlinear equations with four unknowns each. The obtained systems are solved numerically and the results are illustrated in Figure (\ref{Fig: symmetric and asymmetric branches}) for $r=0.76$, $\mu=0.1$ and $\tilde{g}=0.422$.
The results of the numerical simulations reveal the presence of two parabolic branches situated around the computed SIM from equation (\ref{eq: SIM}) obtained via the MSM. These parabolas represent periodic solutions with symmetric impacts. Notably, the minima of these parabolas occur at the same amplitude level $\tilde{C}_{\min}$. The corresponding values of the post- and pre-impact velocities, $B^{\pm}_{\min}$, are equidistant from $B_{\min}$ defined in (\ref{eq: B_min C_tilde_min MSM}). Moreover, as the amplitude $\tilde{C}$ increases, additional solution branches emerge from the parabolas at the same level of excitation, denoted by $\tilde{C}_{\max}$. These branches, exhibiting similar behavior, correspond to periodic motions with two asymmetric impacts per period. The bifurcation point at which the asymmetric solution branches emerge corresponds to a symmetry-breaking bifurcation. A detailed bifurcation analysis leading to the analytical expression for the relevant coordinates $(B^{\pm}_{\max}, \tilde{C}_{\max})$ will be presented in Section \ref{subsection: Linear stabilty and Bif. analysis}.\\ 
To only consider symmetric $\mathcal{P}_2^1$-orbits, the system of equations from (\ref{eq: G +-}) and (\ref{eq: periodicity condition}) can be further simplified by imposing a symmetry condition. This condition states that the first impact of the sequence occurs at $\tau_{n+1}=\tau_n+\frac{\pi}{\Omega}$, and that the absolute value of the post- or pre-impact velocity remains constant but changes only in sign. This symmetry requirement can be reformulated as 
\begin{align}
\begin{pmatrix} \Psi^*_{n+1} \\ B_{n+1}^{\pm *} \end{pmatrix}=
\begin{pmatrix} \Psi^*_n \\ -B_n^{\pm *} \end{pmatrix} +	\begin{pmatrix}\pi \\ 0 \end{pmatrix} \, .
\label{eq: symmetry condition}
\end{align} 
Imposing the symmetry condition (\ref{eq: symmetry condition}) on the discrete mappings  $\mathbf{G}^\pm$, reduces the system of equations for each case to only two equations with two unknowns $\Psi^*=\Psi_n$ and $B^{\pm *}=B_n^{\pm}$ that read:
\begin{align}
\mathbf{G}^+\left(\mathbf{x}^{+*}_n,\, \mathbf{x}^{+*}_{n+2}\right)=\mathbf{0} \Rightarrow \, \left\{
\begin{array}{ll}
C \cos \Psi^* &= R (B^{+*} - \tilde{h})+\tilde{h}\\
C \sin \Psi^* &= 1-\frac{\pi}{2} \left(B^{+*}-\tilde{h}\right)
\end{array} \right. \\
\mathbf{G}^-\left(\mathbf{x}^{-*}_n,\, \mathbf{x}^{-*}_{n+2}\right)=\mathbf{0} \Rightarrow \, \left\{
\begin{array}{ll}
C \cos \Psi^* &= R (B^{-*} + \tilde{h})+\tilde{h}\\
C \sin \Psi^* &= 1-\frac{\pi}{2} \left(B^{-*}+\tilde{h}\right)
\end{array}\right. 
\end{align}
with $\tilde{h}=h(1+\epsilon)$ and the parameters $R$ and $h$ are defined according to (\ref{eq: SIM}). Using the trigonometric identity yields the  following polynomial expressions that relate the post- and pre-impact velocities to the squared value of the external excitation amplitude during the symmetric $2$-IPP motion 
\begin{align}
\tilde{C}&=\left(R (B^{+*} - \tilde{h})+\tilde{h}\right)^2+\left(1-\frac{\pi}{2} \left(B^{+*}-\tilde{h}\right)\right)^2 \label{eq: SIM +} \, ,\\
\tilde{C}&=\left(R (B^{-*} + \tilde{h})+\tilde{h}\right)^2+\left(1-\frac{\pi}{2} \left(B^{-*}+\tilde{h}\right)\right)^2 \label{eq: SIM -}\, .
\end{align}
An illustration of the obtained equations~(\ref{eq: SIM +}) and (\ref{eq: SIM -}) is shown in  Figure~(\ref{Fig: symmetric branches and decaying flow}) together with the SIM~(\ref{eq: SIM}) derived using the MSM in Section~\ref{section: MSM}. Figure~(\ref{Fig: symmetric branches and decaying flow}) also depicts a projection of the system's free resonant response in the~${(B, \, \tilde{C})}$-plane, where~$B$ is the absolute velocity as defined in~(\ref{eq: w ansatz}), and~$B^+$ and~$B^-$ represent the post- and pre-impact velocities, respectively. A closer examination of the derived expressions and numerical simulations, reveals that the slow parameter~$B$, introduced in the MSM framework, corresponds to the average absolute velocity of the absorber and is related to both its post- and pre-impact velocities by
\begin{align}
B=\frac{1}{2}\left(B^+ + B^-\right) \quad \Rightarrow  \quad B_{\min}^\pm= B_{\min}\pm \tilde{h} \, . \label{eq: B average}
\end{align}
Hence, an expression for the SIM for symmetric $\mathcal{P}_2^1$-orbits can be derived based on (\ref{eq: B average}) as 
\begin{align}
\tilde{C}=\left(R B+\tilde{h}\right)^2+\left(1-\frac{\pi}{2} B\right)^2= \underbrace{\left(R B+h\right)^2+\left(1-\frac{\pi}{2} B\right)^2}_{= \text{Eq.} (\ref{eq: SIM}) }+\mathcal{O}\left(\epsilon\right) \label{eq: SIM G} \, ,
\end{align}
This expression reveals that the slow invariant manifold obtained from the impact map approach agrees with the SIM expression derived via the MSM framework up to first order in $\epsilon.$
Notably, the term $\mathcal{O}\left(\epsilon\right)$ highlights higher-order corrections that may introduce slight deviations between the two approaches, yet these deviations remain negligible for $\epsilon \ll 1$.
The analytical expression for $B_{\max}$, which characterizes the symmetry-breaking bifurcation point and the transition to asymmetric orbits, is determined by investigating the stability properties of symmetric $\mathcal{P}_2^1$ orbits. This analysis will be carried out in the next subsection.

\subsection{Linear stability and bifurcation analysis}\label{subsection: Linear stabilty and Bif. analysis}
\begin{figure}[H]
	\centering
	\begin{subfigure}{0.45\textwidth}
		\includegraphics[width=\linewidth]{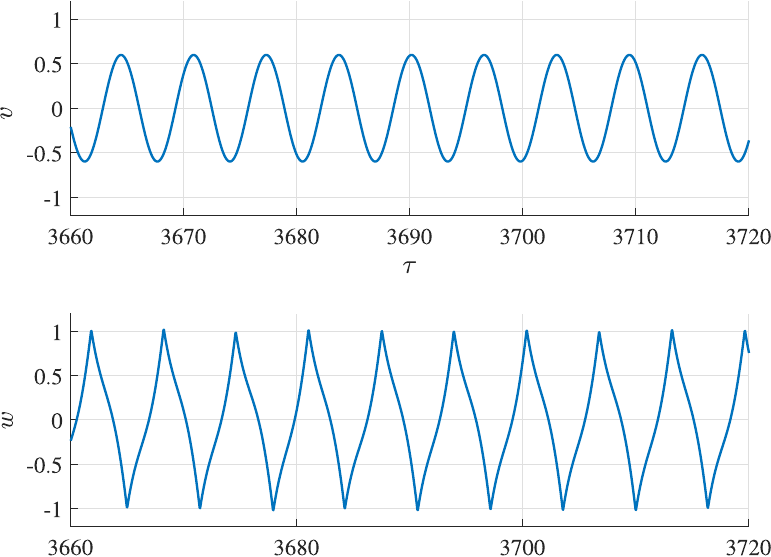}
		\caption{$\Omega=0.977$}
		\label{Fig: verif_response_2_symm_ipp}
	\end{subfigure}\quad
	\begin{subfigure}{0.45\textwidth}
		\includegraphics[width=\linewidth]{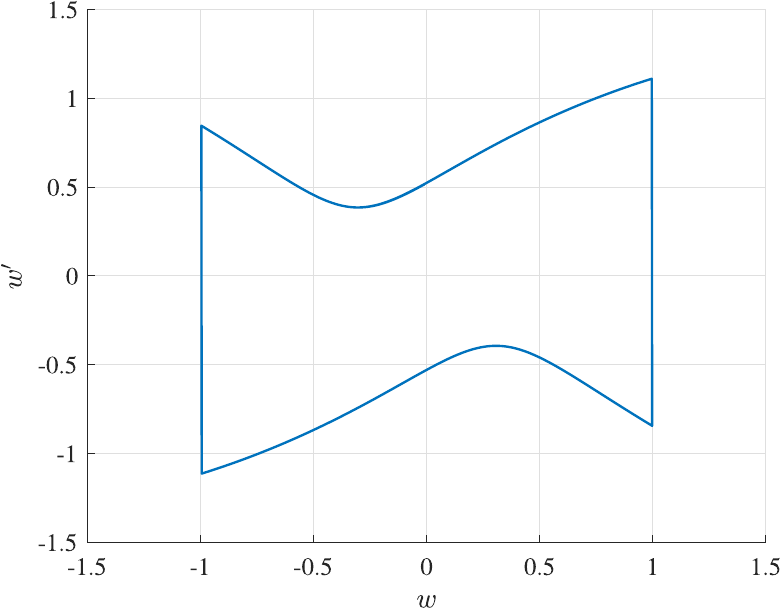}
		\caption{$\Omega=0.977$}
		\label{Fig: verif_PP_2_symm_ipp}	
	\end{subfigure}\\
	\begin{subfigure}{0.45\textwidth}
		\includegraphics[width=\linewidth]{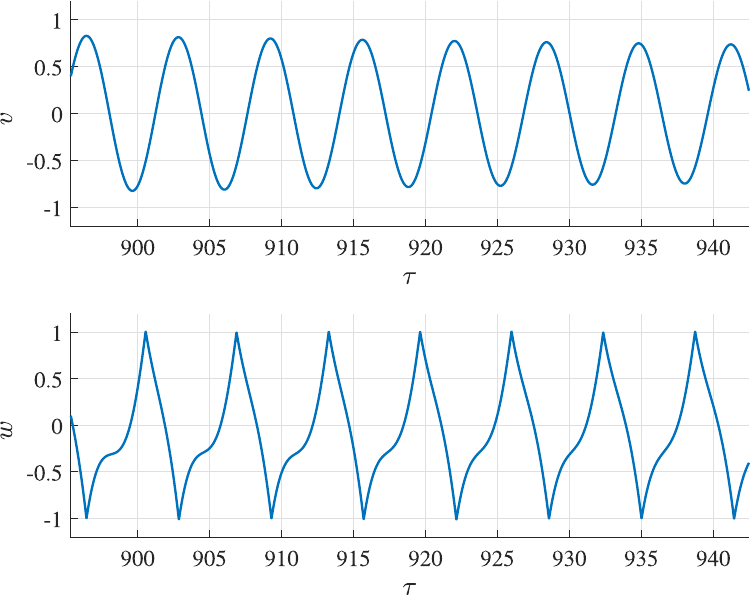}
		\caption{$\Omega=0.983$}
		\label{Fig: verif_response_2_asymm_ipp}
	\end{subfigure}\quad
	\begin{subfigure}{0.45\textwidth}
		\includegraphics[width=\linewidth]{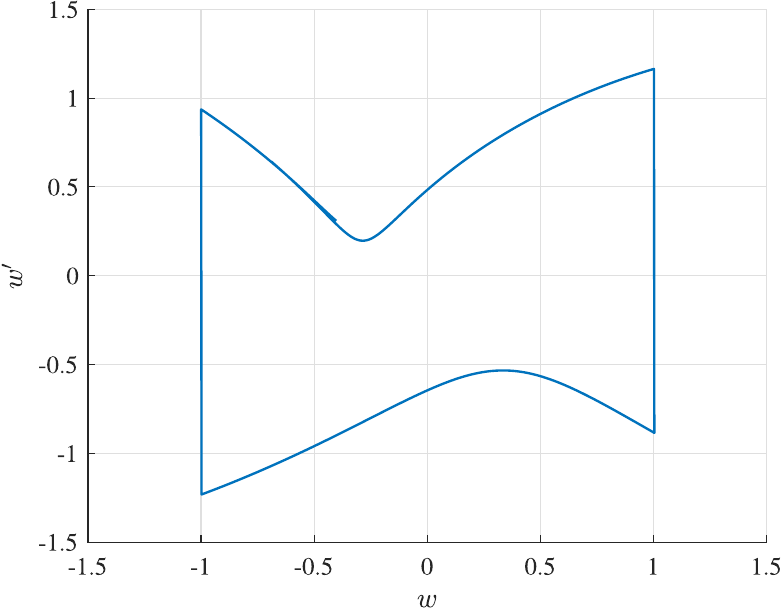}
		\caption{$\Omega=0.983$}
		\label{Fig: verif_PP_2_asymm_ipp}	
	\end{subfigure}\\
	\begin{subfigure}{0.45\textwidth}
		\includegraphics[width=\linewidth]{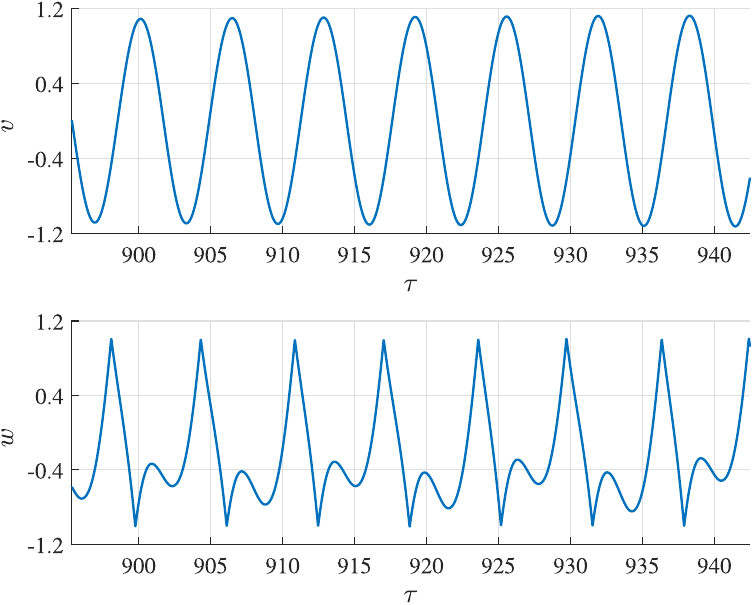}
		\caption{$\Omega=0.986$}
		\label{Fig: verif_response_PD}
	\end{subfigure}\quad
	\begin{subfigure}{0.47\textwidth}
		\includegraphics[width=\linewidth]{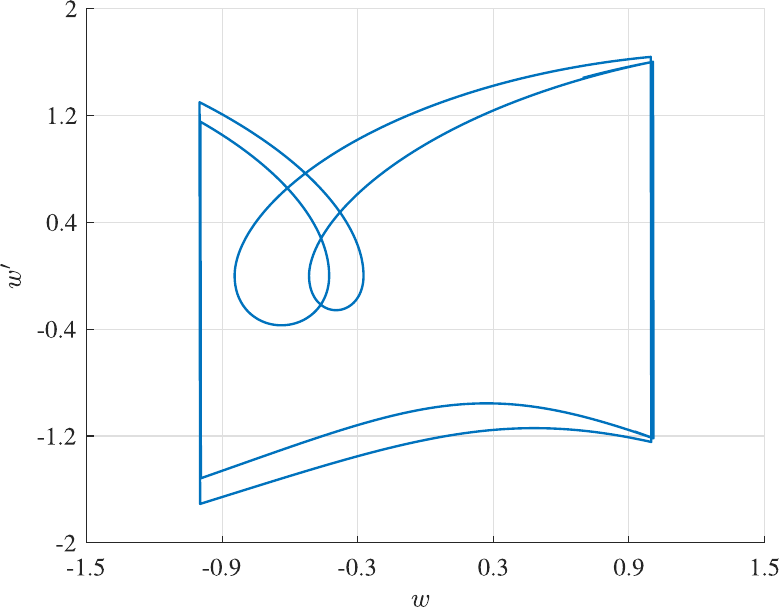}
		\caption{$\Omega=0.986$}
		\label{Fig: verif_PP_PD}	
	\end{subfigure}
	\caption{System response and phase portraits of the absorber periodic motion for the external forcing level $G=3.3$ and the parameters $r=0.76$, $\mu=0.5$ and $\tilde{g}=0.211$: (\ref{Fig: verif_response_2_symm_ipp}, \ref{Fig: verif_PP_2_symm_ipp}) symmetric two impacts per cycle; (\ref{Fig: verif_response_2_asymm_ipp}, \ref{Fig: verif_PP_2_asymm_ipp}) asymmetric two impacts per cycle; (\ref{Fig: verif_response_PD}, \ref{Fig: verif_PP_PD}) four impacts per two cycles.}
	\label{Fig: verification}
\end{figure}

The stability of the $\mathcal{P}_2^1$ orbits can be determined by analyzing the linearized impact map as in \cite{leine2012global, heiman1987dynamics}. This involves introducing small perturbations around the periodic solutions and studying their corresponding propagation through an eigenvalue analysis. The following derived equations and relations in this subsection hold for either choice of the post- or pre-impact state vector, along with the corresponding impact map. To maintain clarity and convenience, the superscripts $\left(\cdot\right)^\pm$ will be omitted. Given that the derivations are intricate, dummy variables are introduced throughout the process to ensure that the steps are easy to follow.\\
The new perturbed state vectors are defined as
\begin{align}
\mathbf{x}_{n+i}=\mathbf{x}^*_{n+i}+\Delta \mathbf{x}_{n+i} \, , \quad \| \Delta \mathbf{x}_{n+i}\|\ll1 \, , \quad i=0,1,2 \, .
\end{align} 
The perturbed solutions $\mathbf{x}_{n+i}$ are by definition also solutions of the system and therefore fulfill~(\ref{eq: G +-}), whereas the fixed points are given by (\ref{eq: periodicity condition}) and (\ref{eq: symmetry condition}). Equation (\ref{eq: G +-}) is expanded as
\begin{align}
\begin{split}
\mathbf{0} &=	\mathbf{G}\left(\mathbf{x}_n,\, 	\mathbf{x}_{n+2}\right) =	\begin{pmatrix}
\mathbf{G}_{\text{LR}}\left(	\mathbf{x}_n,\, 	\mathbf{x}_{n+1}\right) \\
\mathbf{G}_{\text{RL}}\left(	\mathbf{x}_{n+1},\, 	\mathbf{x}_{n+2}\right) \\
\end{pmatrix}\\
&=      
\begin{pmatrix}
\mathbf{G}_{\text{LR}}\left(	\mathbf{x}^*_n,\, 	\mathbf{x}^*_{n+1}\right) +\frac{\partial\mathbf{G}_{\text{LR}}}{\partial \mathbf{x}_{n}}\left(\mathbf{x}^*_n,\, \mathbf{x}^*_{n+1}\right) 	\Delta \mathbf{x}_{n}\cdots \\ \cdots +\frac{\partial\mathbf{G}_{\text{LR}}}{\partial \mathbf{x}_{n+1}}\left(\mathbf{x}^*_n,\, \mathbf{x}^*_{n+1}\right) 	\Delta \mathbf{x}_{n+1} + \cdots \\\\
\mathbf{G}_{\text{RL}}\left(	\mathbf{x}^*_{n+1},\, 	\mathbf{x}^*_{n+2}\right) +\frac{\partial\mathbf{G}_{\text{RL}}}{\partial \mathbf{x}_{n+1}}\left(\mathbf{x}^*_{n+1},\, \mathbf{x}^*_{n+2}\right) 	\Delta \mathbf{x}_{n+1} \cdots \\ \cdots  +\frac{\partial\mathbf{G}_{\text{RL}}}{\partial \mathbf{x}_{n+2}}\left(\mathbf{x}^*_{n+1},\, \mathbf{x}^*_{n+2}\right) 	\Delta \mathbf{x}_{n+2} + \cdots
\end{pmatrix}
\, .
\end{split}
\label{eq: G  perturbed solutions}
\end{align}   
Hence, the propagation of the perturbations can be described through two separate mappings, according to 
\begin{align}
\Delta \mathbf{x}_{n+1} &= - \left(\frac{\partial \mathbf{G}_{\text{LR}}}{\partial \mathbf{x}_{n+1}}\left(\mathbf{x}^*_n,\, \mathbf{x}^*_{n+1}\right)\right)^{-1} \frac{\partial\mathbf{G}_{\text{LR}}}{\partial \mathbf{x}_{n}}\left(\mathbf{x}^*_n,\, \mathbf{x}^*_{n+1}\right) 	\Delta \mathbf{x}_{n} \nonumber\\
& :=\mathbf{A}_{\text{LR}}	\Delta \mathbf{x}_{n} \, , \label{eq: perturbation first impact}\\
\Delta \mathbf{x}_{n+2} &= - \left(\frac{\partial \mathbf{G}_{\text{RL}}}{\partial \mathbf{x}_{n+2}}\left(\mathbf{x}^*_{n+1},\, \mathbf{x}^*_{n+2}\right)\right)^{-1} \frac{\partial\mathbf{G}_{\text{RL}}}{\partial \mathbf{x}_{n+1}}\left(\mathbf{x}^*_{n+1},\, \mathbf{x}^*_{n+2}\right) 	\Delta \mathbf{x}_{n+1} \nonumber\\
&:=\mathbf{A}_{\text{RL}}	\Delta \mathbf{x}_{n+1} \, ,\label{eq: perturbation second impact}
\end{align}
which combined yield a linear mapping from the initial perturbation $	\Delta \mathbf{x}_{n}$ to the perturbation after one period and two impacts $	\Delta \mathbf{x}_{n+2}$ that reads
\begin{align}
\Delta \mathbf{x}_{n+2}=\mathbf{A}\Delta \mathbf{x}_{n} =\mathbf{A}_{\text{RL}} \mathbf{A}_{\text{LR}} \Delta \mathbf{x}_{n} \, . \label{eq: linearized system G}
\end{align}
The linear asymptotic stability of the linearized system (\ref{eq: linearized system G}) is determined by examining the position of the eigenvalues of $\mathbf{A}$ w.r.t.\ the unit circle, i.e., if the eigenvalues of $\mathbf{A}$ satisfy~$|\lambda_{i}|<1$, $i=1,2$, then the fixed point $ \mathbf{x}^*_n$ is asymptotically stable. The eigenvalues of $\mathbf{A}$ can be calculated by solving the equation
\begin{align}
p_{\mathbf{A}}(\lambda)=\lambda^2-\text{Tr}(\mathbf{A}) \lambda+\det(\mathbf{A})=0 \, , \label{eq: char. Polynom}
\end{align}
where $\text{Tr}(\mathbf{A})$ and $\det(\mathbf{A})$ are the trace and determinant of the matrix $\mathbf{A}$, respectively. Determining the roots of the characteristic polynomial (\ref{eq: char. Polynom}) requires the analytical expression of the linearization matrix $\mathbf{A}$ evaluated at a periodic symmetric solution. To achieve this, the conditions of periodicity~(\ref{eq: periodicity condition}) and symmetry~(\ref{eq: symmetry condition}) are inserted into the partial derivatives of the elementary mappings required in (\ref{eq: perturbation first impact}) and (\ref{eq: perturbation second impact}).
For brevity, only the final expression for $\mathbf{A}$ is given here in compact form as
\begin{align}
\mathbf{A}_{\text{RL}}= \begin{pmatrix}
a      & -b \\
-c       & d
\end{pmatrix} \, , \mathbf{A}_{\text{LR}}= \begin{pmatrix}
a      & b \\
c       & d
\end{pmatrix} \, \Rightarrow \quad \mathbf{A}= \begin{pmatrix}
a^2-bc      & b(a-d)\\
c(d-a)       & d^2-bc
\end{pmatrix} \, .
\end{align}
The elements of $\mathbf{A}$ are expressed using the variables $a$, $b$, $c$ and $d$, which are defined as
\begin{align}
\begin{split}
a&=\pi \tilde{k}-\frac{1-R}{1+R} \, ,  \\
b&=\tilde{k}\left(2\tilde{h}-(1-R)(B^\pm\mp \tilde{h})\right)-\frac{1-R}{1+R}\frac{2}{\pi} \tilde{h} \, , \\
c &= \frac{-\pi}{\left(1+R\right)\left(B^\pm\mp \tilde{h}\right)} \, ,\\
d&=\frac{1-R}{1+R}-\frac{2\tilde{h}}{\left(1+R\right)\left(B^\pm\mp \tilde{h}\right)} \, ,
\\
\text{with} \quad \tilde{k}&=\frac{2}{(1+R)^2}\left(\frac{\pi}{2}-\frac{1}{B^\pm\mp \tilde{h}}\right)-\frac{(1-R)\frac{2}{\pi} \tilde{h}}{\left(1+R\right)^2\left(B^\pm\mp \tilde{h}\right)}\, ,
\end{split}
\label{eq: dummy variables}
\end{align}
introduced to simplify the derivations and improve readability.\\
Consequently, using these definitions, the determinant $\det(\mathbf{A})$ and the trace $\text{Tr}(\mathbf{A})$ read
\begin{align}
\det(\mathbf{A}) &=\left(ad-bc\right)^2=r^4 \, , \label{eq: det(A)}\\
\text{Tr}(\mathbf{A})&=\left(a-d\right)^2-2r^2 \,.\label{eq: Tr(A)}
\end{align}
where $r$ is the coefficient of restitution, as previously introduced.
Substituting the expressions for the determinant (\ref{eq: det(A)}) and the trace (\ref{eq: Tr(A)}) into the characteristic polynomial $p_{\mathbf{A}}(\lambda)$ after expressing the eigenvalues in polar coordinates in $\mathbb{C}$ as $\lambda=\rho e^{i\phi}$, with $\rho\geq0$ representing the radial distance to the origin in the complex plane, yields a complex polynomial. Separating the real and imaginary parts of the polynomial and setting them to zero  leads to the following equations
\begin{align}
\operatorname{Re}\left(p_{\mathbf{A}}(\lambda)\right)	&=\rho^2 \cos(2\phi)-	\text{Tr}(\mathbf{A})\rho \cos(\phi)+\det(\mathbf{A})=0 \, , \label{eq: real part of p(lambda)}\\
\operatorname{Im}\left(p_{\mathbf{A}}(\lambda)\right)&=	\rho^2\sin(2\phi)-	\text{Tr}(\mathbf{A})\rho \sin(\phi) =0 \,.\label{eq: imaginary part of p(lambda)}
\end{align}
Solving (\ref{eq: imaginary part of p(lambda)}) leads to two possible cases:
\begin{align}
&\textbf{Case 1:} \quad	 \text{Tr}(\mathbf{A})=2\rho\cos(\phi )\label{eq: Case 1 eig analysis} \,,\\	
&\textbf{Case 2:} \quad \sin(\phi)=0 \, .\label{eq: Case 2 eig analysis}
\end{align}
Each of these cases corresponds to distinct eigenvalue characteristics: Case 1 indicates complex-conjugate eigenvalues, whereas Case 2 corresponds to purely real eigenvalues. To establish the stability criteria, both cases will be systematically analyzed by substituting their respective conditions into the real part of $p_{\mathbf{A}}(\lambda)$ and evaluating the resulting expressions.
\paragraph{$\bullet$ Case 1:  $\lambda_{1,2}\in \mathbb{C}$:\\}
Inserting (\ref{eq: Case 1 eig analysis}) in (\ref{eq: real part of p(lambda)}) yields 
\begin{align}
\operatorname{Re}\left(p_{\mathbf{A}}(\lambda)\right)=0 \quad \Rightarrow \quad	\det(\mathbf{A})=\rho^2\, , \quad \rho\geq0 \, .
\end{align}
Based on the previous derivations, specifically (\ref{eq: det(A)}), the following condition holds
\begin{align}
\rho=r^2 <1 \, , \quad \text{for} \quad 0<r<1 \, . 
\end{align}
Therefore, any complex conjugate pair $\lambda_{1,2}$ is situated within the unit circle, i.e., $|\lambda_{1,2}|=r^2<1$, ensuring the linear asymptotic stability of the corresponding solution.
\paragraph{$\bullet$ Case 2:  $\lambda_{1,2}\in \mathbb{R}$:\\ }
If $\sin(\phi)=0$, then both eigenvalues are situated on the real axis. In this case, instability can arise only if one of the eigenvalues crosses the value $+1$ or $-1$. 
Taking this into consideration, along with stability requirement $\rho<1$, equation (\ref{eq: real part of p(lambda)}) delivers the condition
\begin{align}
\text{Tr}(\mathbf{A})^2<\left(1+\det(\mathbf{A})\right)^2 \label{eq: Trace A condition}
\end{align}
which can be rewritten in terms of the post- or pre-impact absolute velocity $B^\pm$, using (\ref{eq: dummy variables})-(\ref{eq: Tr(A)}) as
\begin{align}
\begin{split}
B^{\pm}_{\min} < B^\pm < B^{\pm}_{\max} & \\
\text{with} \quad  B^{\pm}_{\min}=\frac{2\pi-4R\tilde{h}}{\pi^2+4R^2}\pm \tilde{h} \quad \text{and} \quad B^{\pm}_{\max}&=\frac{2\pi-4R\tilde{h}}{\pi^2-4}\pm \tilde{h}\, .
\end{split}  \label{eq: B bounds G}
\end{align}
Hence, the expressions for the stability bounds $B_{\min}$ and $B_{\max}$, defining the stable branch of the SIM from Equation (\ref{eq: SIM G}), are given by:
\begin{align}
B_{\min}= \frac{2\pi-4R\tilde{h}}{\pi^2+4R^2} \, ,\quad  B_{\max}=\frac{2\pi-4R\tilde{h}}{\pi^2-4} \, .\label{eq: B bounds sim}
\end{align}
These boundaries can be further expressed in terms of the squared excitation amplitude $\tilde{C}$ using Equation (\ref{eq: SIM G}) to define the amplitude range for which the periodic solutions with two symmetric impacts per period are stable, such as
\begin{align}
\tilde{C}_{\min} < \tilde{C} < \tilde{C}_{\max} \, ,\label{eq: C bounds }
\end{align}
with
\begin{align}
\begin{split}
\tilde{C}_{\min}\left(R,\, \tilde{h}\right)&= \frac{\left(R+\frac{\pi}{2}\tilde{h}\right)^2}{\frac{\pi^2}{4}+R^2}=\underbrace{\frac{\left(R+\frac{\pi}{2}h\right)^2}{\frac{\pi^2}{4}+R^2}}_{\text{Eq.}(\ref{eq: B_min C_tilde_min MSM})}+\mathcal{O}\left(\epsilon\right) \,, \\
\tilde{C}_{\max}\left(R, \, \tilde{h}\right)&=\frac{\left(2R\tilde{h}\pi-4\right)^2}{\left(\pi^2-4\right)^2}+\left(\frac{2R(\pi-2R\tilde{h})}{\pi^2-4}+\tilde{h}\right)^2 \, .
\end{split}
\label{eq: Cbounds}
\end{align}
The expression for the activation threshold $\tilde{C}_{\min}$ reveals its dependence not only on the restitution coefficient $r$ (through $R$), the friction coefficient $\mu$ and the cavity length $b$ (through $\tilde{h}$), as in the MSM approach, but also on the mass ratio $\epsilon$. Furthermore, local asymptotic stability of symmetric $\mathcal{P}_2^1$- orbits is guaranteed only for excitation amplitudes within the range $(\tilde{C}_{\min},\tilde{C}_{\max})$, where the average absolute velocity of the absorber satisfies the inequality $B_{\min} < B < B_{\max}$, with the inequality bounds from (\ref{eq: B bounds sim}).
If the amplitude $\tilde{C}$, considered here as the changing bifurcation parameter, goes beyond these bounds, the fixed points lose their stability, potentially leading to the emergence of new solutions. The nature of the new motion depends on the type of bifurcation that occurs, which is determined by how the eigenvalues leave the unit circle. According to (\ref{eq: det(A)}), the eigenvalues cannot leave the unit circle as a complex conjugate pair, since their product has been shown to satisfy $\lambda_1 \lambda_2 =r^4 <1$. Therefore, bifurcations occur only for real-valued eigenvalues. If an eigenvalue of the linearized map evaluated at the fixed point moves through $+1$, then one of the following bifurcations can occur: first, a saddle-node bifurcation, characterized by the simultaneous creation or destruction of stable and unstable solution branches. The corresponding bifurcation point represents a boundary for the existence of solutions in the parameter space. The second type of bifurcation is a symmetry-breaking bifurcation, characterized by the creation of asymmetric solutions from symmetric ones.  When an eigenvalue is at $-1$, only a period-doubling bifurcation occurs. In this case, the new steady-state motion has twice the period of the one it emerges from. In this context, solving $p_{\mathbf{A}}(\lambda=+1)=0$ results in the same values presented in~(\ref{eq: B bounds G}) and subsequently~(\ref{eq: Cbounds}). This means that the bifurcations that occur on the SIM are either a saddle-node, or a symmetry-breaking bifurcation. However, period-doubling bifurcations do not occur on the SIM, since $p_{\mathbf{A}}(\lambda=-1)=0$ has no solutions. 
\begin{figure}[t!]
	\centering
	\begin{minipage}{0.475\textwidth}
		\includegraphics[width=\linewidth]{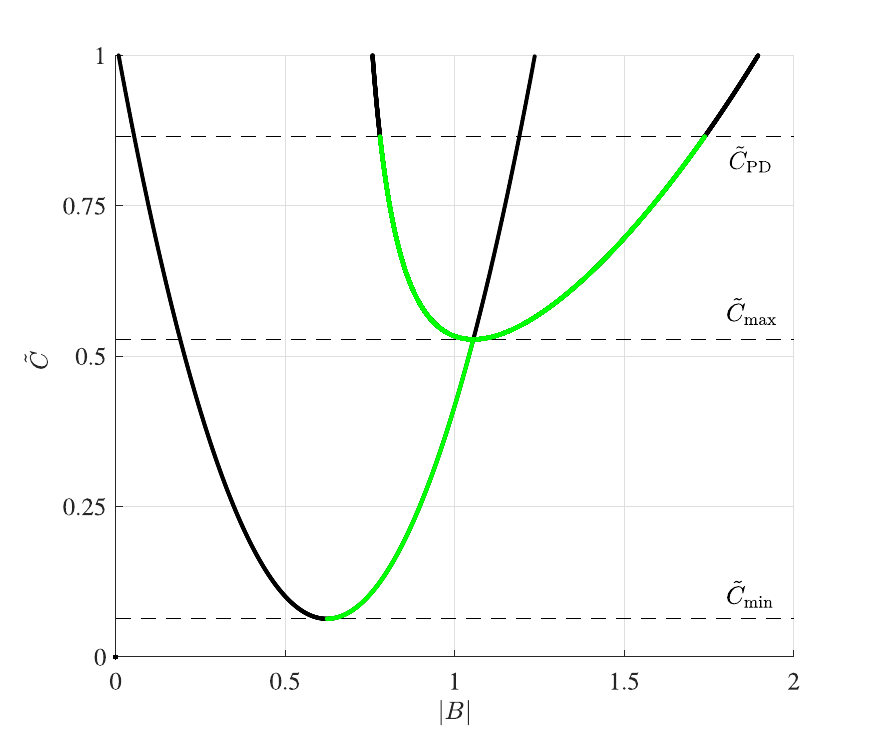}
		\caption{Stability region of the $\mathcal{P}_2^1$-orbits for $r=0.76$, $\mu=0.5$ and $\tilde{g}=0.211$. The black lines result from $\mathbf{G}^{\pm}\left(\mathbf{x}_n,\, \mathbf{x}_{n+2}\right)=\mathbf{0}$ and (\ref{eq: B average}). The green lines correspond to the stable branches.}
		\label{Fig: stability region of 2ipp}
	\end{minipage}\quad
	\begin{minipage}{0.475\textwidth}
		\includegraphics[width=\linewidth]{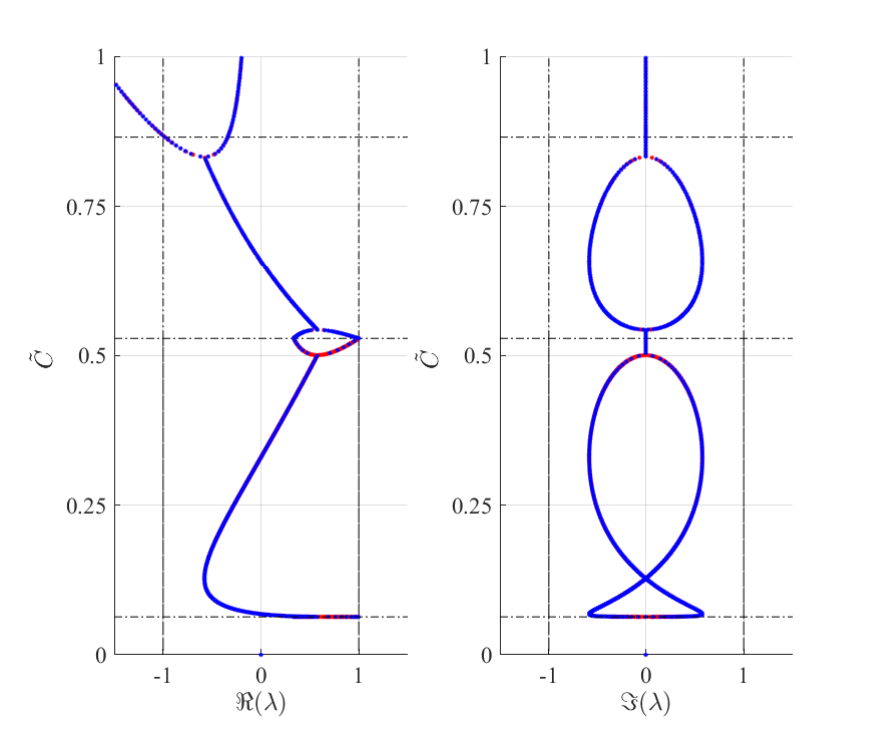}
		\caption{Eigenvalues of the matrices $\mathbf{A}^\pm$ for $r=0.76$, $\mu=0.5$ and $\tilde{g}=0.211$.\\The red and blue lines corresponding to the post- and pre-impact cases, respectively, overlap perfectly. }
		\label{Fig: eigenvalues 2ipp}	
	\end{minipage}
\end{figure}
The numerical results shown in Figure~(\ref{Fig: stability region of 2ipp}) are consistent with the analytical findings. The computed solution branches of the $\mathcal{P}_2^1$ orbits are obtained through solving the nonlinear equations resulting from $\mathbf{G}^\pm\left(\mathbf{x}_n, \mathbf{x}_{n+2}\right)=\mathbf{0}$ along with the periodicity condition (\ref{eq: periodicity condition}) and (\ref{eq: B average}). The stable branches are determined through an eigenvalue analysis of the numerically computed matrices $\mathbf{A}^\pm$.  The computed solution branches present three types of bifurcations corresponding to the points where the stability property of each solution branch is lost. The behavior of the eigenvalues along these branches is illustrated in Figure (\ref{Fig: eigenvalues 2ipp}), where the real and imaginary parts of the eigenvalues are displayed as functions of the parameter $\tilde{C}$. The most immediate observation from Figure (\ref{Fig: eigenvalues 2ipp}), is that the evolution of the eigenvalues with increasing value of the amplitude $\tilde{C}$ generated from the numerical solutions of both equations $\mathbf{G}^\pm\left(\mathbf{x}_n, \mathbf{x}_{n+2}\right)=\mathbf{0}$ coincide exactly as expected, since $\mathbf{G}^+$ and $\mathbf{G}^-$ describe the same motion in different coordinates. This provides an accurate prediction for the bifurcating behavior of the average solution $B$ from (\ref{eq: B average}). Moreover, it is clear that $\tilde{C}_{\min}$ corresponds to an eigenvalue passing through $+1$, and it represents the point from which two solution branches emerge — only one of which is stable. As $\tilde{C}$ increases, the eigenvalues remain within the unit circle until $\tilde{C}_{\max}$ is reached. The stable branch between $\tilde{C}_{\min}$ and $\tilde{C}_{\max}$ corresponds to stable symmetric $\mathcal{P}_2^1$-orbits. At the critical value of $\tilde{C}_{\max}$, for which the corresponding eigenvalue passes through $+1$ again, the symmetry property is lost, and the stable asymmetric $\mathcal{P}_2^1$-orbits emerge from the SIM. The stability of the new branches is maintained as $\tilde{C}$ increases until a period-doubling bifurcation point is reached at $\tilde{C}= \tilde{C}_{\text{PD}}$ when an eigenvalue passes through $-1$.\\
So far, the analysis has been conducted by assuming a constant $\Omega$. However, the expressions of the impact maps reveal a dependency of the solutions not only on the amplitude $\tilde{C}$ but also on the normalized frequency~$\Omega$. In the following, the sensitivity of the system's behavior to the choice of $\Omega$ is highlighted, particularly in the vicinity of $1$, i.e., $\omega\approx\omega_0$. For a better illustration of the system behavior along the different branches of the $\mathcal{P}_2^1$-orbits for different values of $\Omega$, the displacement of the absorber within the cavity as well as the corresponding phase-portraits at steady state are shown in Figure~(\ref{Fig: verification}). The system parameters and the level of external forcing, i.e.\ $G$, are kept constant during the simulations. The position of the steady-state solution has been influenced by the change in the frequency of the external excitation. For $\Omega=0.977$, the steady-state amplitude of the main structure $\tilde{C}$ is within the interval $\left(\tilde{C}_{\min}, \, \tilde{C}_{\max}\right)$, which corresponds to a symmetric motion with only two impacts per excitation cycle as depicted in Figure (\ref{Fig: verif_response_2_symm_ipp})-(\ref{Fig: verif_PP_2_symm_ipp}). For $\Omega=0.983$, the motion period remains equal to the excitation period, and the impacts are no longer symmetric, which corresponds to $\tilde{C}_{\max}<\tilde{C}<\tilde{C}_{\text{PD}}$ as shown in Figures (\ref{Fig: verif_response_2_asymm_ipp})-(\ref{Fig: verif_PP_2_asymm_ipp}). By choosing $\Omega=0.986$, the response amplitude is above the critical value $\tilde{C}_{\text{PD}}$ and the motion with period $2T$ is observed. This motion is characterized by 4 impacts per two excitation cycles and is illustrated in Figures (\ref{Fig: verif_response_PD})-(\ref{Fig: verif_PP_PD}). It's important to note that the solution branches of such $\mathcal{P}_4^2$-orbits, can be computed numerically using new impact maps defined as: 
\begin{align}
\mathbf{G}\left(\mathbf{x}_n,\, \mathbf{x}_{n+4}\right)=
\begin{pmatrix}
\mathbf{G}_{\text{LR}}\left(\mathbf{x}_n,\, \mathbf{x}_{n+1}\right) \\
\mathbf{G}_{\text{RL}}\left(\mathbf{x}_{n+1},\, \mathbf{x}_{n+2}\right) \\
\mathbf{G}_{\text{LR}}\left(\mathbf{x}_{n+2},\, \mathbf{x}_{n+3}\right) \\
\mathbf{G}_{\text{RL}}\left(\mathbf{x}_{n+3},\, \mathbf{x}_{n+4}\right) \\
\end{pmatrix} =\mathbf{0} \, . \label{eq: G +- pd}
\end{align}   
with the new periodicty condition 
\begin{align}
\mathbf{x}_n^{*}=\begin{pmatrix}	\Psi^*\\	B^{*} \end{pmatrix} \, ,\quad \mathbf{x}_{n+4}^{ *}=\begin{pmatrix}	\Psi^*+4\pi\\	B^{ *} \end{pmatrix} \, .
\end{align}
In summary, the solutions go through a symmetry-breaking bifurcation initially, followed by a cascade of period-doubling bifurcations that leads to chaos.
In this work, the elementary mappings required to describe any impact sequence for a general case of $\mathcal{P}_l^k$-orbits, i.e., periodic motions with $l$ impacts per $k$-cycle, are introduced along with the corresponding stability and bifurcation analysis. The sought solutions result from solving a set of nonlinear algebraic equations built as 
\begin{align}
\mathbf{G}\left(\mathbf{x}_n,\, \mathbf{x}_{n+l}\right)=
\begin{pmatrix}
\mathbf{G}_{m_1}\left(\mathbf{x}_n,\, \mathbf{x}_{n+1}\right) \\
\vdots\\
\mathbf{G}_{m_l}\left(\mathbf{x}_{n+l-1},\, \mathbf{x}_{n+l}\right) \\
\end{pmatrix} =\mathbf{0} \, ,  \label{eq: G general}
\end{align}
with $m_1, \cdots , \, m_l \in \left\{ \text{LR}, \,\text{RL},\, \text{RR}, \, \text{LL}\right\}$ and the periodicity condition 
\begin{align}
\mathbf{x}_n^{ *}=\begin{pmatrix}	\Psi^*\\	B^{*} \end{pmatrix} \, ,\quad \mathbf{x}_{n+l}^{ *}=\begin{pmatrix}	\Psi^*+2k\pi\\	B^{ *} \end{pmatrix} \, . \label{eq: periodicity condition general}
\end{align} 
The explicit analysis of more complex motions and their corresponding bifurcation analysis is not within the scope of this work.
\section{Nonlinear response characteristics}\label{Section: response regimes}
This section discusses the different operating ranges of the impact absorber and the corresponding response regimes, focusing on on the system response under near-resonant forcing. To this end, the methods from Sections \ref{section: MSM}-\ref{section: Impact Map} are exploited to better predict the dynamics of the studied system. Each method offers specific insights into different aspects of the system's dynamics. Assuming a time scales separation leading to slow-fast dynamics, the impact map provides a systematic way to determine periodic solutions and their stability, i.e.\ the system's steady-state behavior. It effectively identifies the set of all possible solutions, particularly those that represent the simplest, yet most effective type of response in terms of dissipation, i.e. the symmetric 2 impacts per period. Additionally, it reveals the local stability properties of these solutions, identifying a set of locally attractive solutions towards which the system may converge.
In constrast, the MSM captures the slow-scale dynamics (with respect to the slow time scale $\tau_1$) along these solution branches. Specifically, it predicts the existence of the steady-state solutions as well as the slow-flow behavior around the equilibrium on the SIM.
Essentially, a comprehensive understanding of the system's behavior is provided by combining both approaches.

\subsection{Effect of friction on design-relevant performance metrics}
This subsection presents closed-form expressions for the key design-relevant performance metrics of the VI-NES, derived from the Multiple Scales Method and the impact-map formulation, focusing on the symmetric~$\mathcal{P}_2^1$-orbits. These metrics quantify how dry friction influences activation, stability, resonance-frequency shift, and decay. The expressions for the activation and amplitude bounds were derived in Sections~\ref{section: MSM}-\ref{section: Impact Map}, while the backbone curve and decay characteristics are obtained here by extending the procedure previously applied to the frictionless case in~\cite{YOUSSEF2021116043}. To maintain conciseness, intermediate algebraic steps are omitted and only the final closed-form expressions, reformulated in terms of the physical system parameters, and their physical implications are discussed.
\paragraph{Amplitude bounds and optimal activity range}
The stability bounds of the symmetric~$\mathcal{P}_2^1$-orbits define the main performance metrics of the VI-NES, corresponding to the amplitude level at which the absorber engages and the upper limit beyond which the optimal response for TET ceases. Reformulated in terms of the absolute displacement of the primary structure using the approximation 
\begin{equation} 
q_1(t) \approx b \sqrt{\tilde{C}(\epsilon \omega_0 t)} \sin(\omega_0 t + \gamma) \, ,
\end{equation}
for a small mass ratio~$\epsilon$, yields the following physical bounds:
\begin{align}
q_{1,\min} &= b \sqrt{\tilde{C}_{\min}}= \frac{\pi\, b \,h}{\sqrt{\pi^2+4R^2}}+\frac{2 \, b R}{\sqrt{\pi^2+4R^2}} \nonumber\\
&=\mu \frac{\pi^2\ g}{2\omega_0^2\sqrt{\pi^2+4R^2}}+b\frac{2   R}{\sqrt{\pi^2+4R^2}} \, ,\label{eq: Physical activation threshold q1_min}\\
q_{1,\max} &= b \sqrt{\tilde{C}_{\max}} \nonumber \\
&= \frac{2\pi b}{\pi^2-4} \sqrt{\left(R\frac{\mu}{b} \frac{g}{\omega_0^2}\frac{\pi}{2}-\frac{2}{\pi}\right)^2 +\left(R-\frac{\mu}{b} \frac{g}{\omega_0^2}\left(R^2+1-\left(\frac{\pi}{2}\right)^2\right)\right)^2} \, . \label{eq: Physical activation threshold q1_max} 
\end{align} 
Here, the use of~$R$ preserves analytical clarity while highlighting the role of~$r$ without unnecessary expansion.
It should be noted that the tuning parameters might affect the normalized and physical amplitudes differently: for instance, increasing~$b$ decreases~$\tilde{C}_{\max}$ while increasing~$q_{1,\max}$. This highlights the importance of basing design considerations on physical rather than normalized quantities.  
When defining the effective operating window of the symmetric~$2$-IPP regime, however, both formulations lead to the same dimensionless measure,
\begin{equation} 
\xi = \frac{C_{\max}}{C_{\min}}
= \sqrt{\frac{\tilde{C}_{\max}}{\tilde{C}_{\min}}}= \sqrt{\frac{\frac{q_{1,\max}^2}{b^2}}{\frac{q_{1,\min}^2}{b^2}}}=\frac{q_{1,\max}}{q_{1,\min}}  \, ,
\label{eq:active_range_2IPP}
\end{equation}
which measures the width of the stable amplitude range and thus characterizes the energetic robustness of the VI-NES. 
\paragraph{Decay rate and effective damping} 
Following~\cite{YOUSSEF2021116043}, the free resonant response of the primary structure coupled to the VI-NES is examined. The amplitude evolution along the SIM is approximated using an upper linear bound of the slow dynamics,
\begin{align}
D_1 \tilde{C} \leq -\tilde{\alpha}_c \tilde{C} + \tilde{K}_c \, ,\label{eq:upper bound of d1C}
\end{align}
where $\tilde{\alpha}_c$ and~$\tilde{K}_c$ are defined in~(\ref{eq: Coefficients}).
Solving~(\ref{eq:upper bound of d1C}) yields the analytical envelope~$\tilde{C}$, used to approximate the absolute displacement of the primary structure as
\begin{equation}
q_{1,\text{app}} (t) =  \sqrt{ \left(\left(\frac{q_{10}}{\sin\gamma}\right)^2 - \frac{\tilde{K}_c b^2}{\tilde{\alpha}_c} \right) e^{-\tilde{\alpha}_c\epsilon \omega_0 t} +\frac{\tilde{K}_c b^2}{\tilde{\alpha}_c} } \sin\left(\omega_0 t +\gamma\right) \, ,\label{eq: q1 app, large c} 
\end{equation}
with~${q_{10}:=b\sqrt{\tilde{C}(t=0)} \sin\gamma}$. Here, the decay rate~$\tilde{\alpha}_c$ as well as the amplitude of the exponential term jointly characterize the overall dissipation strength.
For the frictionless case studied in~\cite{YOUSSEF2021116043}, the impact-induced damping ratio~$D_{\text{NES}} \left(R, \, \epsilon\right)$, defined therein, provides a baseline dissipation level.
In the presence of friction, the total dissipation increases as larger~$\mu$ produces steeper decay and faster energy reduction. However, enhanced dissipation comes at the cost of a higher activation threshold, a trade-off that will be discussed in the following section.
\paragraph{Resonance shift and backbone curve}
The nonlinear resonance properties of the coupled LO-VI-NES system are characterized through the backbone curve, derived by applying the phase-resonance condition within the slow-flow framework, following the procedure established in~\cite{YOUSSEF2021116043}. 
Substituting the condition~$\gamma_{R} = -\frac{\pi}{2} + 2 j \pi$, $j \in \mathbb{Z}$, into the steady-state equations obtained by setting the right-hand side of~(\ref{eq: slow dynamic impacting}) to zero yields the steady-state amplitude–forcing and detuning relations,
\begin{align}
G\sqrt{\tilde{C}_{R}}&= \tilde{\alpha}_{c} \tilde{C}_{R}+ \tilde{\beta}_{c} \sqrt{\tilde{C}_{R}-\tilde{C}_{\text{min}}}-\tilde{K}_{c}  \label{eq: C=f(G) at ss} \, ,\\
0&= \tilde{\alpha}_{\gamma} \tilde{C}_{R}+ \tilde{\beta}_{\gamma} \sqrt{\tilde{C}_{R}-\tilde{C}_{\text{min}}}-\tilde{K}_{\gamma}  \label{eq: sigma=f(C) at ss} \, ,
\end{align} 
with coefficients defined in~(\ref{eq: Coefficients}). Here, the subscript $(\cdot)_R$ denotes quantities evaluated at resonance and should not be confused with the restitution-related parameter~$R = \frac{1-r}{1+r}$.
Rearranging~(\ref{eq: sigma=f(C) at ss}) and reformulating in physical coordinates gives the resonance frequency
\begin{align}
\begin{split}
f_R&=f_0(1+\epsilon \sigma_R)\\
&=f_0\left(1+ \frac{\epsilon}{2b\, q_{1,R}} \left(-  \tilde{\beta}_{\gamma}  \sqrt{q_{1,R}-q_{1,\min}} +\tilde{K}_{\gamma}  \right)-\frac{\epsilon}{\frac{\pi^2}{4}+R^2}\right)
\end{split}\, ,\label{eq: resonance frequency shift}
\end{align}
where the coefficients~$\tilde{\beta}_{\gamma}$, $\tilde{K}_{\gamma}$, and~$q_{1,\min}$ depend on~$\mu$, $b$, and~$R$ (see~(\ref{eq: Coefficients})). 
Equation~(\ref{eq: resonance frequency shift}) defines the backbone curve of the LO-VI-NES system, which connects the maxima of the frequency-response curves obtained for different excitation levels~\cite{peter2016tracking}.  
Here, a compact representation is preferred over a full expansion, as it retains parameter dependencies and preserves analytical transparency for design purposes.
\paragraph{Discussion}
The analytical results provide quantitative insight into how dry friction affects the performance of the VI-NES within the optimal operating regime for targeted energy transfer.
The influence of friction is neither inherently beneficial nor detrimental but depends on the excitation conditions and desired performance objectives. Increasing friction raises the activation threshold, leading to delayed engagement under harmonic excitation or earlier disengagement during free decay. This behavior is advantageous for harmonic excitation, ensuring VI-NES activation near resonance, but can limit energy dissipation during free decay, where a low activation threshold is preferable. Conversely, a low activation threshold can trigger early activation under harmonic loading, leading to higher-order periodic responses dominated by impacts and reduced frictional contribution near resonance.\\
While friction enhances dissipation, it simultaneously reduces the VI-NES activity bandwidth by narrowing both the frequency range and amplitude interval of efficient TET. \\
Overall, the influence of friction is strongly amplitude dependent: within the~$2$-IPP regime, it introduces competing effects that lead to design trade-offs, whereas at higher amplitudes, its effect becomes negligible as the response converges toward the frictionless limit~\cite{youssef2025asymmetric}.

\subsection{Response regimes}
In the frictionless case, when the only nonlinearity introduced in the system arises from impacts between the absorber and the main structure, i.e., $\mu=0$, the system can exhibit three different types of response. At low levels of excitation, the activation condition of the absorber is generally not met, causing both masses to oscillate in a primarily linear fashion. In this regime, the absorber remains inactive while the structure behaves as a linear oscillator without the influence of the auxiliary attachment, although random impacts between the masses might still occur.
At higher levels of excitation that activate the VI-NES, i.e., when the amplitude $\tilde{C}$ exceeds the threshold $\tilde{C}_{\min}$ and the VI-NES average velocity is above $B_{\min}$, a periodic response with constant amplitude is observed. The steady-state solution corresponds to a fixed point of the impact map~$\mathbf{G}\left(\mathbf{x}_n,\,\mathbf{x}_{n+l}\right)=\mathbf{0}$ as defined in (\ref{eq: G general})-(\ref{eq: periodicity condition general}), for $\mu=0$. This type of solution is characterized by $l$ impacts per $k$-cycles. The third response type, known in the literature as (chaotic) strongly modulated response (SMR), arises in the absence of stable steady-state solutions or when only unstable ones exist. This type of response is hard to predict. During this regime, the system's flow is intermittently attracted to the slow invariant manifold, where it remains on it for certain intervals, corresponding to phases with sustained two impacts per period and a reduction of the main structure’s amplitude. Once the amplitude reaches the activation threshold, i.e.\ $\tilde{C}_{\min}$, the flow leaves the SIM, causing both masses to oscillate independently, as in the first case. As the amplitude $\tilde{C}$ increases due to external forcing, the flow could land in the vicinity of the attractive branch of the SIM and is drawn back to it. These cycles are repeated in a non-periodic fashion, leading to the chaotic nature of the response.\\
In the presence of dry friction, the observed response regimes are slightly different. While the activation condition of the absorber might appear similar, i.e., $\tilde{C}\geq \tilde{C}_{\min}$ and $B\geq B_{\min}$, dry friction significantly affects the system's behavior. In the following, the different response regimes are presented along with the simulation results. 
\subsubsection{Inactive absorber: Linear response }
\begin{figure}[t!]
	\centering
	\begin{minipage}{0.575\textwidth}
		\includegraphics[width=\linewidth]{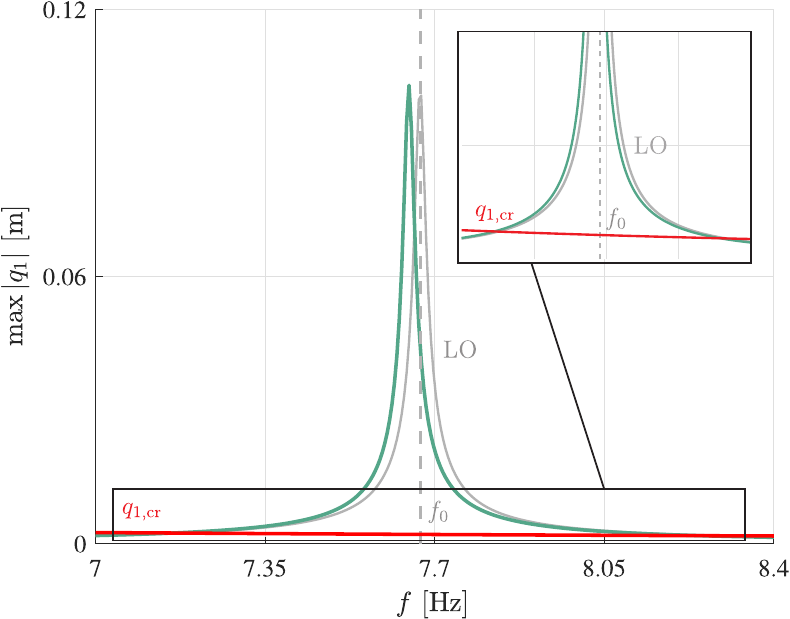}	
	\end{minipage}
	\caption{Frequency response curve of the linear oscillator without any attachment (grey line) for the external level of excitation $G=2.47$. The blue line corresponds to the critical values $q_{1, \text{cr}}$ from (\ref{eq: q_1,cr}) and the dashed line corresponds to the periodic response regime with a fully sticking auxiliary mass.}		
	\label{Fig: q1_cr}	
\end{figure}
In this scenario, the auxiliary mass is sticking to the lower surface of the cavity without sliding, even as the primary mass oscillates. Even though the system behaves linearly, it is worth noting that the resonance frequency is slightly shifted since the effective mass of the system is slightly increased by the mass of the absorber. For the absorber to stick permanently to the surface of the cavity, the friction force $\lambda_T$, as defined in (\ref{eq: Coulomb's law}), must be large enough to counteract the inertial force caused by the oscillations of the main structure. This condition can be reformulated into the following inequality
\begin{align}
(1+\epsilon) \mu \tilde{g} \geq C\Omega^2  \quad \Leftrightarrow \quad  C \leq (1+\epsilon)\mu \tilde{g}  \frac{1}{\Omega^2}:= C_{\text{cr}} \, , \label{eq: sticking limit} 
\end{align}
where the parameters $\Omega$ and $\tilde{g}$ are defined as in (\ref{eq: parameter normalisation}). 
The critical value $C_{\text{cr}}$ depends not only on the normalized frequency $\Omega$ and the friction coefficient $\mu$, but also on the mass ratio $\epsilon$ and the cavity length $b$ (through $\tilde{g}$). The frequency response curve (FRC) in Figure (\ref{Fig: q1_cr}) represents the maximum amplitude of the system's response, i.e., $|q_1|_{\max}$, as a function of the frequency of the external excitation $f=\frac{\omega}{2\pi}$, for the linear oscillator with a sticking auxiliary mass.
In this context, the critical displacement values of the main structure, $q_{1,\text{cr}}$, can be approximated according to (\ref{eq: normalization}) as
\begin{equation}
q_1=b \left(\frac{1}{1+\epsilon}v+\frac{\epsilon}{1+\epsilon}w\right) \quad \Rightarrow \quad q_{1,\text{cr}} \approx \frac{b}{1+\epsilon} C_{\text{cr}}=\mu g \frac{1}{\omega^2}\, . \label{eq: q_1,cr}
\end{equation}
Hence, the dependency on $\epsilon$ and $b$ is effectively omitted, and the critical values defining the existence range for this response regime are mainly dependent on the friction coefficient $\mu$. Moreover, from Figure (\ref{Fig: q1_cr}), it is clear that the amplitude levels corresponding to this response regime are very low. The solution branches within this range nearly coincide with those of the linear case, and therefore, this regime will not be considered in the prediction of the system's behavior near resonance.

\subsubsection{Purely sliding absorber: almost linear response}
Another scenario involves the absorber sliding back and forth within the cavity without coming in contact with the side walls. This type of motion is directly related to whether the absorber's velocity within the cavity satisfies its activation condition, as well as the initial conditions of the system, specifically where the flow starts in the slow variables space $\left(B,\, \tilde{C}\right)$. Although this case may initially seem similar to the frictionless scenario where $\mu=0$, it reveals one of the major effects of dry friction on the behavior of the system at hand. Specifically, during this motion, the auxiliary mass acts as a sliding absorber, with an additional energy dissipation due to friction. Moreover, the presence of friction generates an additional manifold within the slow variable space. The method of multiple scales can be employed again to predict the existence of equilibria on this manifold, following the same procedure as outlined previously.\\
In this scenario, the acceleration and relative velocity of the absorber within the cavity are described by 
\begin{align}
D^2_0 w_0(\tau_0, \tau_1)&=D^2_0 v_0(\tau_0,\tau_1)-\mu\tilde{g}M\left(\tau_0+\eta(\tau_1)\right) \, ,\\
D_0 w_0(\tau_0, \tau_1)&=D_0 v_0(\tau_0, \tau_1)-\mu \tilde{g}\Pi\left(\tau_0+\eta(\tau_1)\right) \, .\label{eq: rel velocity sliding absorber}
\end{align}
In other words, this case corresponds to setting $B=0$ in (\ref{eq: w0}) and (\ref{eq: w ansatz}). 
According to (\ref{eq: rel velocity sliding absorber}), the time evolution of the normalized absolute velocity of the auxiliary mass during this particular type of motion, denoted by~$\tilde{q}_2^\prime \approx w_0^\prime - v_0^\prime$, oscillates between $\pm h=\pm \frac{\pi}{2}\mu \tilde{g}$. 
Therefore, the expression for the relative displacement of the absorber within the cavity, $w_0$, is given by
\begin{align}
w_0(\tau_0, \tau_1)&= 	v_0(\tau_0, \tau_1)  -\frac{1}{2} \mu \tilde{g} \left(\Pi^2\left(\tau_0+\eta(\tau_1)\right)-\frac{\pi^2}{4}\right) M\left(\tau_0+\eta(\tau_1)\right) \,.\label{eq: w ansatz: sliding absorber}
\end{align}
Here, the constant value of $B=0$ defines a new vertical manifold in the $(B,\tilde{C})$ parameter space. However, this manifold exists only within a limited amplitude range for $\tilde{C}$, as the sliding motion is constrained by the sidewalls of the cavity.
\paragraph{Different motion scenarios:}
Within this sliding regime, two extreme cases can be distinguished: the first is where the auxiliary mass is moving back and forth with a very small displacement. This case of minimal sliding will be referred to as an "almost sticking" absorber. The second case involves larger displacements of the auxiliary mass, still within the cavity, where it may come close to, or even grazes, the sidewalls without generating impulsive dynamics. This will be described as "almost impacting" absorber. Both scenarios can be described by the same ansatz from (\ref{eq: w ansatz: sliding absorber}),  where the time instant at which the motion of the auxiliary mass is reversed, i.e., when its velocity changes sign, is denoted by $\tau_{0,k}^{r}$. The motion reversal condition for these cases can be formulated as follows:
\begin{align}
\left|w_0(\tau^r_{0,k})\right|=\delta  \, , \quad 	D_0^+ w_0(\tau^r_{0,k}) = D_0^- w_0 (\tau^r_{0,k})=D_0 w_0(\tau^r_{0,k})=0 \, .\label{eq:motion reversal condition}
\end{align}
Here, the parameter $\delta\in \left(0,1\right)$ describes the maximal displacement of the mass within the cavity, with $\delta \rightarrow 0$ corresponding to an "almost sticking" absorber and  $\delta \rightarrow 1$ corresponding to an "almost impacting" absorber.

\paragraph{Symmetric motion and multiple scales analysis:}
It is important to note that the ansatz~(\ref{eq: w ansatz: sliding absorber}) describes a symmetric periodic motion of the mass, where the time needed for the mass to move from the left to the right is equal to the time needed to move back. This symmetry condition is reflected in the choice of the time instant $\tau_{0,k}^{r}$ defined accordingly as $\tau_{0,k}^{r}=\frac{\pi}{2}+k\pi-\eta$. The method of multiple scales is employed again to derive the slow dynamics of the amplitude and phase during the sliding motion of the auxiliary mass, resulting in the following equations: 
\begin{align}
\begin{split}
D_1 C&= -\frac{1}{2} \left(\lambda C - \frac{16-\pi^4}{\pi^4} h \cos\left(\eta - \theta \right) - G \sin \left(\sigma\tau_1-\theta\right)\right)  \\
D_1 \theta &= -\frac{1}{2 C} \left(-\frac{16-\pi^4}{\pi^4} h \sin\left(\eta - \theta \right) + G \cos \left(\sigma\tau_1-\theta\right)\right)
\end{split}
\label{eq: slow dynamics sliding}
\end{align} 
Depending on the considered motion and the chosen $\delta$, the corresponding ODE system describing the slow dynamics along the slow invariant manifold ($B=0$) can be derived using the condition from (\ref{eq:motion reversal condition}). Accordingly, the sine and cosine terms in (\ref{eq: slow dynamics sliding}) can be expressed as 
\begin{align}
\left|w_0(\tau^r_{0,k})\right|=\delta 	\quad &\Rightarrow C \cos\left(\eta-\theta\right)=\delta  \, , \label{eq: cos sliding absorber}\\
D_0 w_0 (\tau^r_{0,k}) =0	\quad & \Rightarrow C \sin\left(\eta-\theta\right)=\frac{\pi}{2}\mu \tilde{g}=h   \label{eq: sin sliding absorber}\, . \\
\quad\text{For} \quad \delta\rightarrow 0 \, : \quad \Rightarrow \cos\left(\eta_\text{sl}-\theta\right)&=0 \, , \quad  \sin\left(\eta_\text{sl}-\theta\right)=\frac{h}{C} \, .\label{eq: rev condition almost impacting absorber} \\
\quad\text{For} \quad \delta\rightarrow 1 \, : \quad \Rightarrow \cos\left(\eta_\text{sl}-\theta\right)&=\frac{1}{C} \, , \quad \sin\left(\eta_\text{sl}-\theta\right)=\frac{h}{C} \, .\label{eq: rev condition almost sticking absorber} 
\end{align}
Applying the trigonometric identity, the maximum displacement of the absorber within the cavity, $\delta$, is expressed in terms of the squared amplitude $\tilde{C}$ as
\begin{equation}
\delta=\sqrt{\tilde{C}-h^2}\quad \text{with} \quad \tilde{C}\geq h^2\, .\label{eq: delta C_tilde relation}
\end{equation} 
Using (\ref{eq: delta C_tilde relation}), the slow dynamics along the newly introduced manifold are derived as in Section \ref{section: MSM}, resulting in a set of equations that describe the slow evolution of the squared amplitude $\tilde{C}$ and the phase $\gamma$ along this manifold w.r.t.\ the slow time scale $\tau_1$. These equations are given by
\begin{align}
\begin{split}
D_1 \tilde{C}&=-\lambda \tilde{C}+\frac{16-\pi^4}{\pi^4} h \sqrt{\tilde{C}-h^2}-G\sqrt{\tilde{C}} \sin \left(\gamma\right) \, ,\\
D_1 \gamma &=-\frac{1}{2\tilde{C}} \left(G\sqrt{\tilde{C}} \cos\left(\gamma\right)-\frac{16-\pi^4}{\pi^4} h^2+2\sigma \tilde{C}\right) \,.
\end{split}\label{eq: slow dynamic sliding absorber}
\end{align}
\paragraph{Operational range:}
To account for all possible sliding motions within the interval $\delta \in \left(0, \, 1\right)$, the extreme cases, i.e., $\delta=0$ and $\delta=1$, are examined. These cases mark the critical transitions from sticking to sliding or from sliding to near-contact with the cavity walls, respectively. 
This analysis provides the amplitude range within which, the auxiliary mass behaves as a purely sliding absorber. 
By substituting the bounds of $\delta$ into (\ref{eq: delta C_tilde relation}), the existence range for this response regime is obtained as
\begin{align}
\tilde{C}_\text{sl, min}:= h^2\leq  \tilde{C}=\delta^2+h^2\leq 1+h^2 =: \tilde{C}_\text{sl, max} \, . \label{eq; sliding absorber bounds}
\end{align}
These amplitude bounds are directly derived from (\ref{eq: delta C_tilde relation}), ensuring their consistency with the slow dynamic equations in (\ref{eq: slow dynamic sliding absorber}).
\paragraph{Steady-state solutions and stability analysis:}
\begin{figure}[t!]
	\centering
	\begin{minipage}{0.575\textwidth}
		\includegraphics[width=\linewidth]{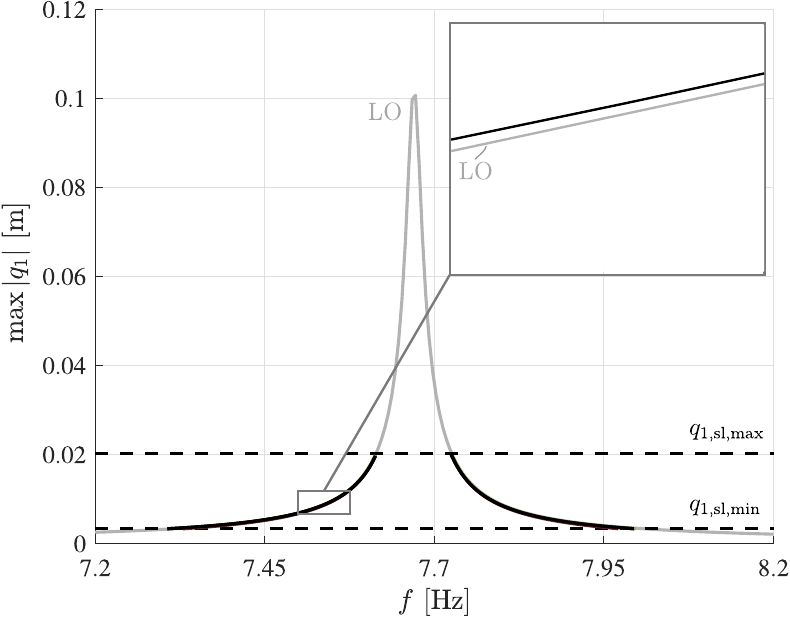}	
	\end{minipage}
	\caption{Approximated frequency response curves of the LO coupled to a purely sliding absorber (black line) for $\mu=0.5$ and $\tilde{g}=0.211$ and the external excitation level $G=2.47$. The grey line corresponds to the LO response without an attachment.}
	\label{Fig: FRC sliding absorber}
\end{figure}
The steady-state solutions along the sliding motion manifold, are determined by finding equilibria of (\ref{eq: slow dynamic sliding absorber}), which are given as the positive real roots of the following polynomial equation
\begin{align}
\left(\lambda^2+4\sigma^2\right)\tilde{C}^2+\left(2\lambda p h \delta+4\sigma p h^2-G^2\right)\tilde{C}+p^2h^2\left(h^2+\delta^2\right)= 0 \,  , \; p=\frac{\pi^4-16}{\pi^4} \, . \label{eq: SS sliding absorber}
\end{align}
By substituting (\ref{eq: delta C_tilde relation}), this equation can be rewritten as the following nonlinear algebraic equation:
\begin{align}
\left(\lambda^2+4\sigma^2\right)\tilde{C}+2\lambda p h \sqrt{\tilde{C}-h^2}+4\sigma p h^2-G^2+p^2h^2= 0 \,  , \quad p=\frac{\pi^4-16}{\pi^4} \, . \label{eq: SS sliding absorber2}
\end{align}
The positive real roots of (\ref{eq: SS sliding absorber2}) correspond to the equilibria on the sliding motion manifold, which stability is then assessed through an eigenvalue analysis of the Jacobian matrix associated with the system (\ref{eq: slow dynamic sliding absorber}). The approximated frequency response curves are presented in Figure (\ref{Fig: FRC sliding absorber}). These curves lie close to the linear response curve and exist only within the amplitude range bounded by $q_{1,\text{sl},\max}$ and $q_{1,\text{sl},\min}$ given by
\begin{align}
\begin{split}
q_{1,\text{sl},\min}&=\frac{b}{1+\epsilon}\sqrt{ \tilde{C}_{\text{sl},\min}}=\frac{b}{1+\epsilon}h \, , \\ q_{1,\text{sl},\max}&=\frac{b}{1+\epsilon} \sqrt{\tilde{C}_{\text{sl},\max}}=\frac{b}{1+\epsilon}\sqrt{1+h^2} 
\end{split}
\label{eq: bounds purely sliding absorber}
\end{align}
\begin{figure}[t!]
	\centering
	\begin{minipage}{0.575\textwidth}
		\includegraphics[width=\linewidth]{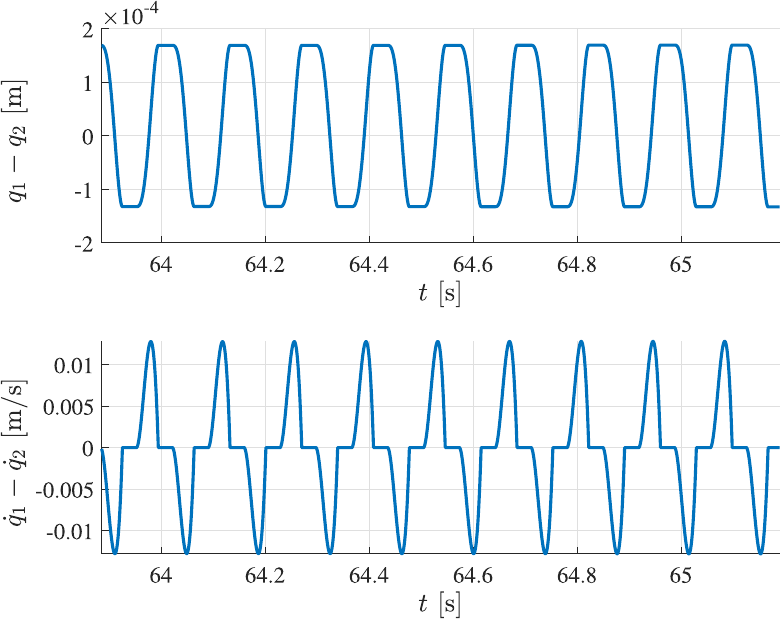}	
	\end{minipage}
	\caption{System response with stick-slip motion at steady state for $r=0.76$, $\mu=0.5$, $\tilde{g}=0.211$ and the external forcing parameters $G=2.47$ and $\sigma=-9.04$ ($f=7.25$ Hz), simulated using (\ref{eq: eq of motion 2a})–(\ref{eq: eq of motion 2b}).}
	\label{Fig: stick-slip}
\end{figure}
It is important to highlight that $q_{1,\text{sl},\min}$ from (\ref{eq: bounds purely sliding absorber}) and $q_{1,\text{cr}}$ from (\ref{eq: sticking limit}) are distinct values and do not represent the same threshold.
Specifically, $q_{1,\text{sl},\min}$ represents the minimal amplitude beyond which purely sliding motion occurs, while $q_{1,\text{cr}}$ marks the amplitude threshold below which the absorber remains permanently sticking to the lower surface of the cavity. The motion within this intermediate range follows a stick-slip pattern, which cannot be captured by the MSM anstaz employed in this study. This behavior is beyond the scope of this paper, as it requires a zero relative velocity during the stick phases, which contradicts the assumptions made in (\ref{eq: MSM assumption}). Figure~(\ref{Fig: stick-slip}) illustrates an example of such a motion showing a stick-slip pattern.

\subsubsection{Active vibro-impact absorber: Constant amplitude response}\label{subsubsection:Active vibro-impact absorber: Constant amplitude response}
\begin{figure}[t!]
	\centering
	\begin{minipage}{0.575\textwidth}
		\includegraphics[width=\linewidth]{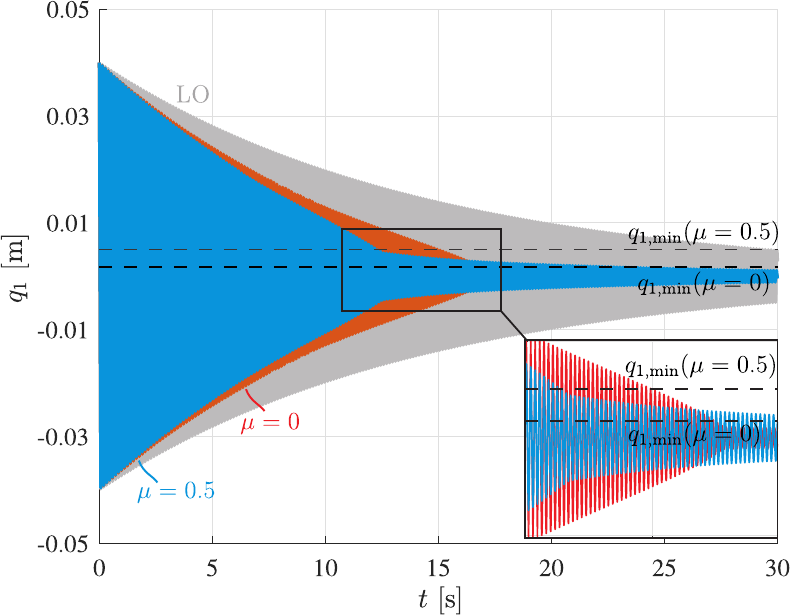}
	\end{minipage}
	\caption{Comparison of the effect of friction on the free vibrations of the main structure. The blue and red lines correspond to $\mu=0.5$ and $\mu=0$ respectively. The system parameters are $r=0.76$ and $\tilde{g}=0.211$. The grey line corresponds to the LO response without an attachment.}	
	\label{Fig: Free vibrations friction vs frictionless}
\end{figure}

Finally, the case of a constant amplitude response with an impacting absorber is investigated. In this regime, the auxiliary mass undergoes periodic impacts with the sidewalls of the cavity, resulting in discrete energy exchanges between the primary structure and the absorber. These impacts introduce an additional dissipation mechanism that modifies the characteristics of the system's response.
This behavior is fundamentally different from the sliding regime, where the absorber moves within the cavity without contact, dissipating the energy continuously through dry friction, and from the sticking regime, where the auxiliary mass sticks to the surface. The transition to an impacting absorber introduces an additional mechanism of energy dissipation through collisions, modifying the effective damping of the system.
\begin{figure}[t!]
	\centering
	\begin{minipage}{0.575\textwidth}
		\includegraphics[width=\linewidth]{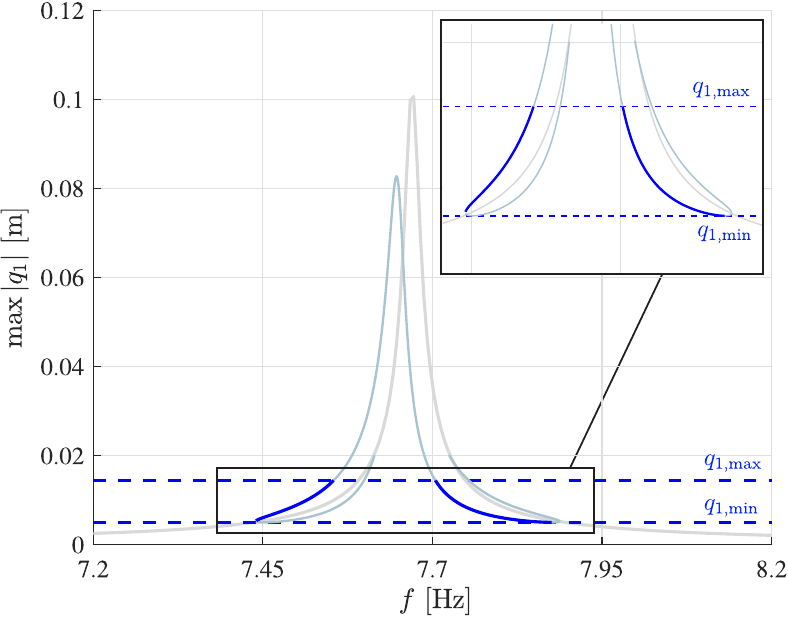}
	\end{minipage}
	\caption{Approximated frequency response curves of the LO coupled to an impacting absorber for $\mu=0.5$ and $\tilde{g}=0.211$ and the external level of excitation $G=2.47$. The dark blue and faded blue lines represent the stable and unstable branches, respectively. The grey line corresponds to the LO response without an attachment.}	
	\label{Fig: FRC impacting absorber}	
\end{figure}
Compared to the frictionless case from \cite{YOUSSEF2021116043}, the constant amplitude response regime with an impacting absorber remains qualitatively similar but with different bounds and existence conditions, mainly influenced by the friction coefficient, as detailed in Sections \ref{section: MSM}-\ref{section: Impact Map}. Figure (\ref{Fig: FRC impacting absorber}) illustrates the stable branches of the approximated frequency response curve, obtained using Equations~(\ref{eq: SIM}) and~(\ref{eq: ESIM}). The active response range of the absorber is determined by the bounds in (\ref{eq: C bounds })-(\ref{eq: Cbounds}). The stability of these branches is verified through an eigenvalue analysis of the Jacobian of the corresponding slow system (\ref{eq: slow dynamics impacting absorber}), which determines whether the equilibria on the attractive branch of the SIM are locally asymptotically stable. It has been established that the presence of friction alters the stability regions and affects the achievable periodic responses. Specifically, it shifts the position of the SIM (\ref{eq: SIM}) in the slow variables space, influencing the system’s behavior and the location of the equilibria.
\begin{figure}[t!]
	\centering
	\begin{subfigure}{0.475\textwidth}
		\includegraphics[width=\linewidth]{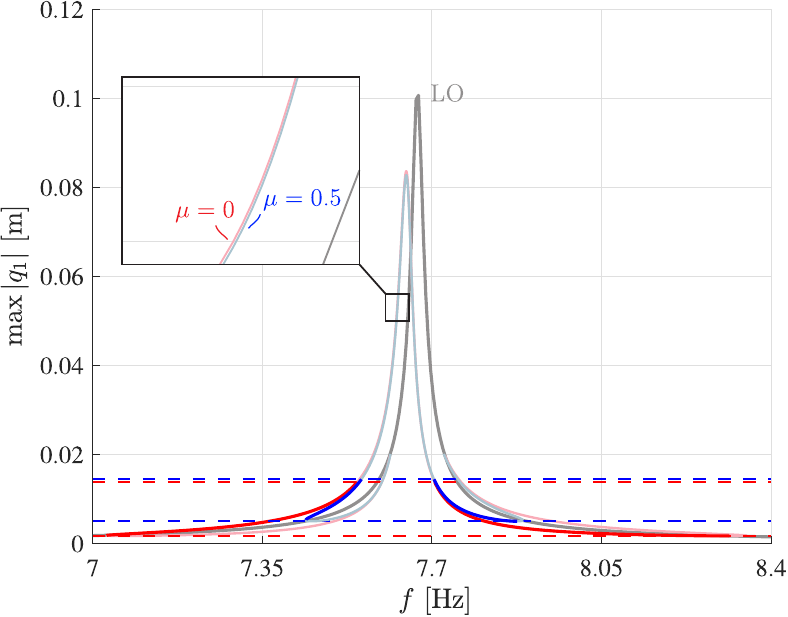}
		\caption{Frequency response curves.}
	\end{subfigure}\quad
	\begin{subfigure}{0.475\textwidth}
		\includegraphics[width=\linewidth]{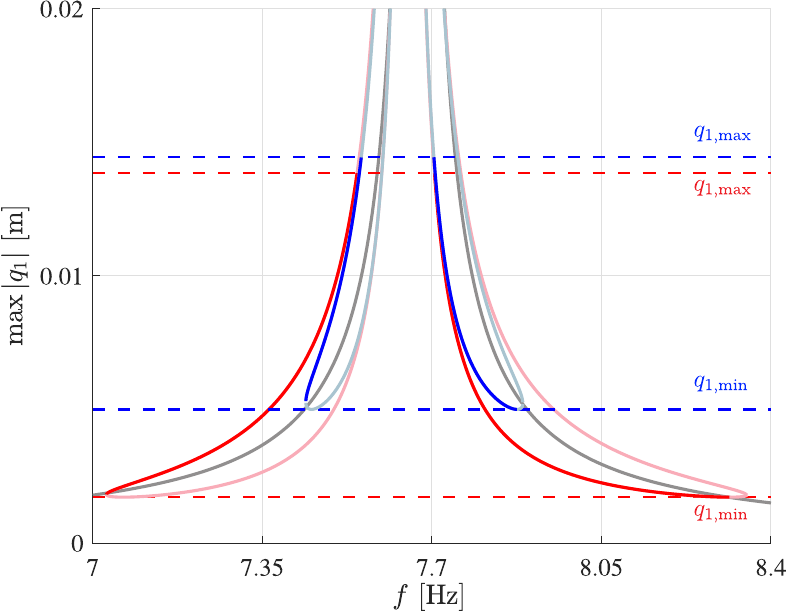}
		\caption{Zoomed-in view.}
	\end{subfigure}
	\caption{ Comparison of the effect of friction on the stable periodic solutions with 2 symmetric impacts per period. The blue and red lines correspond to $\mu=0.5$ and $\mu=0$, respectively. The faded lines correspond to the unstable branches. The system parameters are $r=0.76$ and $\tilde{g}=0.211$ and the external level of excitation is $G=2.47$. The grey line corresponds to the LO response without an attachment.}	
	\label{Fig: friction vs frictionless}
\end{figure}
The influence of friction on the system’s response becomes clearer when comparing the frequency response curves of the frictionless case to those with friction. The main observation is that, in the constant amplitude response regime, friction reduces the active frequency range of the VI-NES and increases the activation threshold, i.e., $\tilde{C}_{\min}$, as shown in Figure~(\ref{Fig: friction vs frictionless}). While these effects may appear unfavorable, they can be effectively controlled and managed during the design process, ensuring effective performance even when the ideal frictionless case cannot be achieved.
Regarding the damping performance of the VI-NES with and without friction, each system offers superior performance in different frequency regions. The frictionless system performs better above the shifted resonance frequency, whereas the system with friction is more effective below it. As the frequency approaches resonance and the amplitude accordingly increases, the influence of friction diminishes, leading both systems to exhibit similar performance as the solution branches of both cases merge.
At low amplitude levels, friction becomes non-negligible. As the amplitude approaches $q_{1,{\min}}$, i.e., $\tilde{C} \rightarrow \tilde{C}_{\min}$, the dissipation rate increases, thereby enhancing the overall damping effect of the VI-NES. This behavior is illustrated in Figure~(\ref{Fig: Free vibrations friction vs frictionless}), where the free vibrations of the main structure are compared for both systems, shown in Figure~(\ref{Fig: friction vs frictionless}). The results indicate that, although friction narrows the absorber’s activation frequency range and raises the activation threshold, it also enhances the decay rate and energy dissipation at lower response levels. This trade-off should be carefully considered in the design process to optimize the performance according to the system's specific requirements.

\subsection{Coexistence of solutions:}
\begin{figure}[t!]
	\centering
	\begin{minipage}{0.675\textwidth}
		\includegraphics[width=\linewidth]{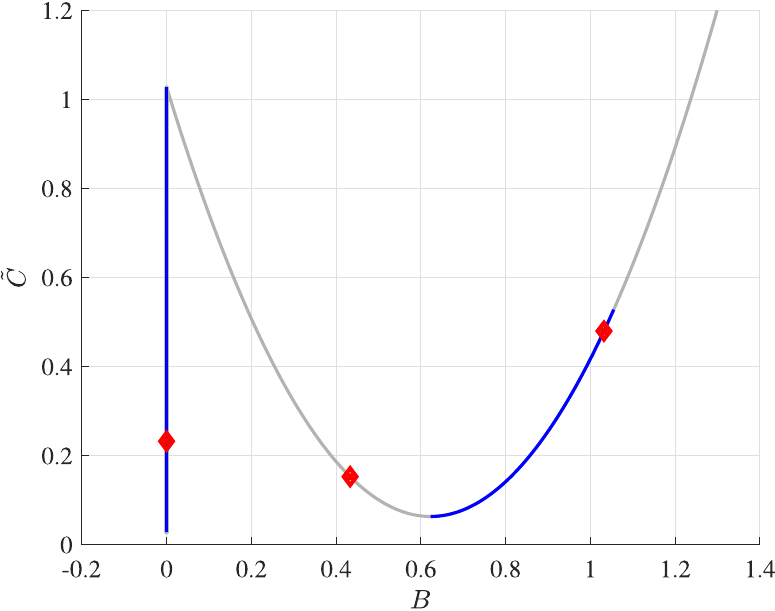}
	\end{minipage}
	\caption{Equilibria of the systems from (\ref{eq: slow dynamic impacting}) and (\ref{eq: slow dynamic sliding absorber}) situated on the corresponding slow manifolds. The equilibria are depicted in red, and the stable branches of the slow manifolds are depicted in blue. The system parameters are $r=0.76$, $\mu=0.5$ and $\tilde{g}=0.211$ and the external excitation parameters are $G=2.47$ and $\sigma=-2.58$.}	
	\label{Fig: coexistence of SS}	
\end{figure}

\begin{figure}[t!]
	\centering
	\begin{subfigure}{0.475\textwidth}
		\includegraphics[width=\linewidth]{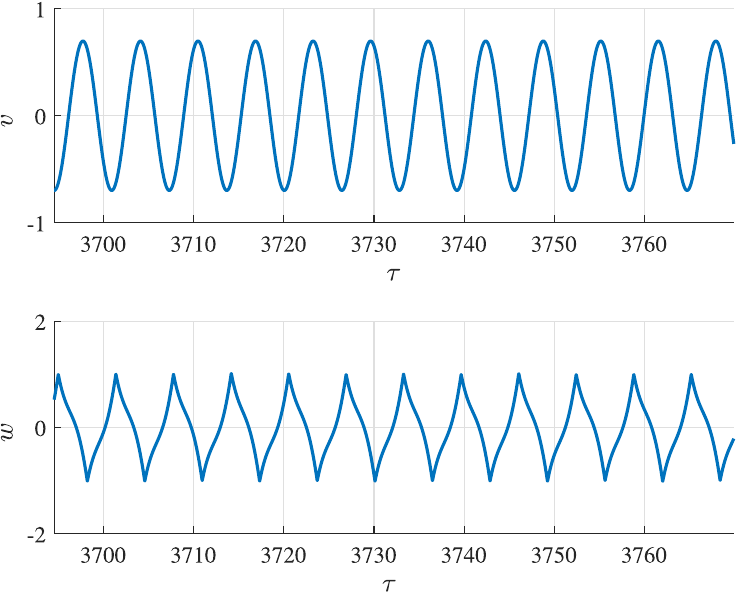}
		\caption{System response at steady state.}
		\label{Fig: impacting_absorber_response}
	\end{subfigure}\quad
	\begin{subfigure}{0.475\textwidth}
		\includegraphics[width=\linewidth]{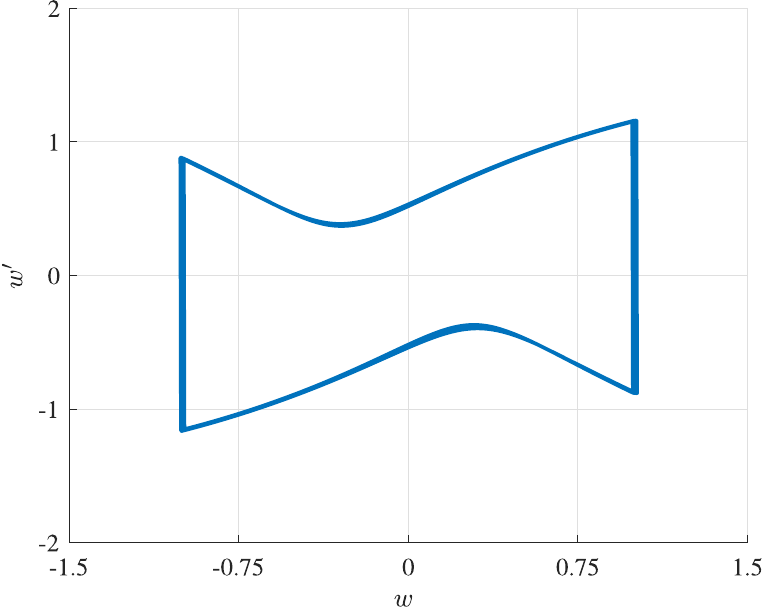}
		\caption{Phase portrait.}
		\label{Fig: impacting_absorber_PP}	
	\end{subfigure}\\
	\begin{subfigure}{0.475\textwidth}
		\includegraphics[width=\linewidth]{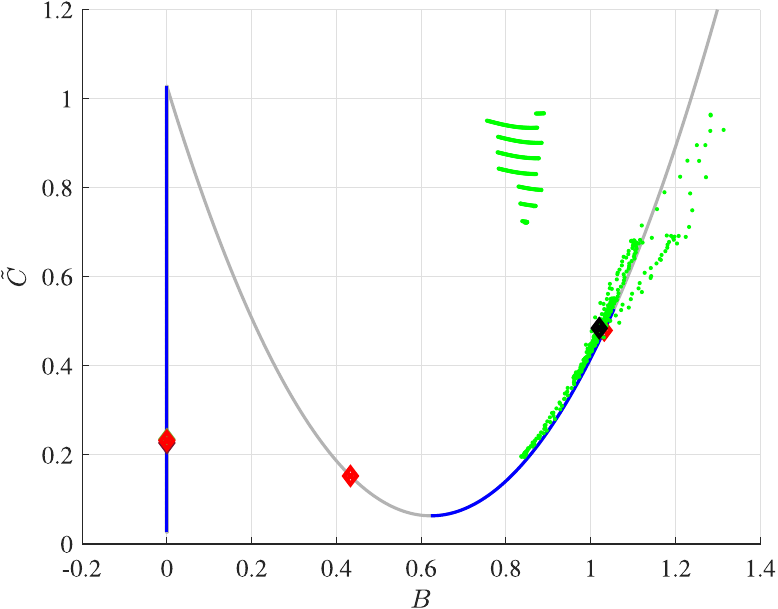}
		\caption{Projection of the flow (green line) on the $(B,\tilde{C})$-plane. The red points correspond to the equilibria of the systems from (\ref{eq: slow dynamic impacting}) and (\ref{eq: slow dynamic sliding absorber}). The black points represents the simulated steady-state amplitude.}
		\label{Fig: impacting_absorber_flow_C_B}
	\end{subfigure}\quad
	\begin{subfigure}{0.475\textwidth}
		\includegraphics[width=\linewidth]{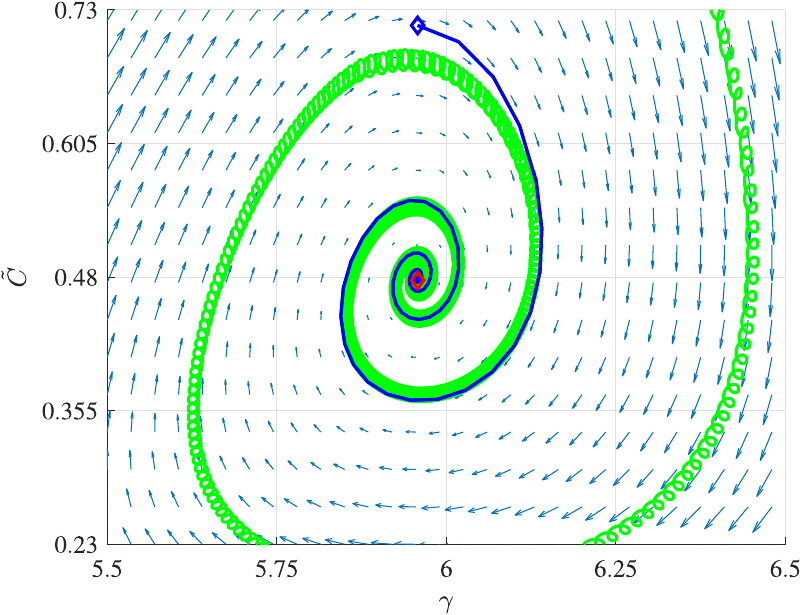}
		\caption{Projection of the flow (green line) on the $(\gamma,\tilde{C})$-plane. The blue line represents the flow of the system governed by (\ref{eq: slow dynamic impacting}) and the red point indicates the corresponding equilibrium.}
		\label{Fig: impacting_absorber_flow_C_gamma}	
	\end{subfigure}
	\caption{System response and phase portraits of the absorber's periodic motion for $r=0.76$, $\mu=0.5$, $\tilde{g}=0.211$ and the external forcing parameters $G=2.47$ and $\sigma=-2.58$,  simulated using (\ref{eq: eq of motion 2a})–(\ref{eq: eq of motion 2b}). For the initial conditions $q_1(0)=q_2(0)=0.02 \left[\text{m}\right]$ and $\dot{q}_1(0)=\dot{q}_2(0)=0 \left[\text{m/s}\right]$ , the auxiliary mass acts as an active vibro-impact absorber.}
	\label{Fig: impacting absorber}
\end{figure}

In the previous subsections, the behavior of the system in different response regimes was examined, revealing the existence of a new vertical manifold ($B=0$) in the slow variables space. This manifold, along with the associated steady-state solutions, can influence the dynamics by potentially limiting the number of SMR cycles. In fact, the solution might converge toward the vicinity of this new manifold, ultimately leading to a periodic solution that closely resembles the linear response. Additionally, for a given input, multiple equilibria may exist on both manifolds, as described by Equations (\ref{eq: slow dynamic impacting}) and (\ref{eq: slow dynamic sliding absorber}), where the equilibria represent the steady-state solutions of the slow dynamics on the respective manifold. An example of coexistent solutions on both manifolds is shown in Figure (\ref{Fig: coexistence of SS}). The red points represent the computed equilibria. Among those situated on the SIM of two symmetric impacts per period, one lies on the attractive branch and is stable, while the other is unstable.
Even though the solution will not converge to the unstable equilibria, multiple stable periodic solutions with distinct domains of attraction may occur. This implies that the choice of initial conditions becomes crucial, as it determines the stable periodic solution to which the system will eventually converge. 
The following time-domain simulation results illustrate the system's response in such a case for two different sets of initial conditions. For both scenarios, the motion is projected onto the $(B,\tilde{C})$-plane and the $(\gamma,\tilde{C})$-plane to monitor the flow's evolution along the respective manifolds. First, the case where the initial conditions are chosen as $q_1(0)=q_2(0)=0 \left[\text{m}\right]$ and $\dot{q}_1(0)=\dot{q}_2(0)=0 \left[\text{m/s}\right]$ is considered. 
\begin{figure}[t!]
	\centering
	\begin{subfigure}{0.475\textwidth}
		\includegraphics[width=\linewidth]{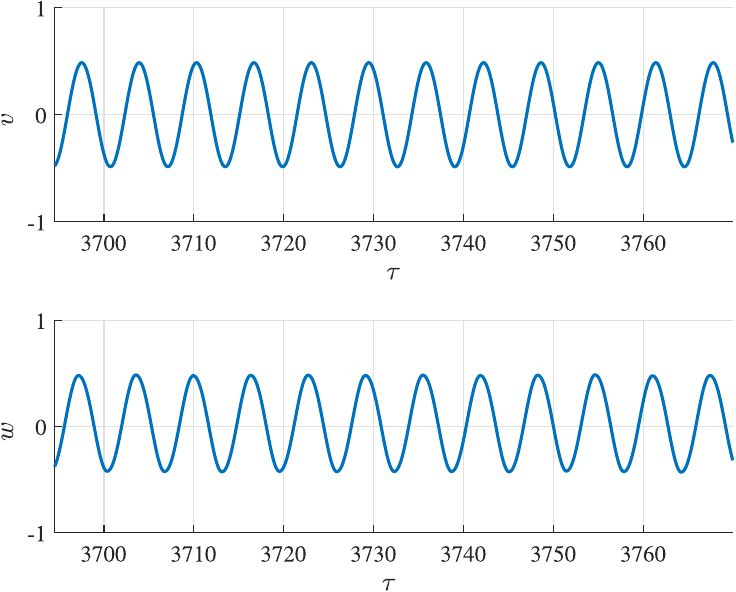}
		\caption{System response at steady state.}
		\label{Fig: sliding_absorber_response}
	\end{subfigure}\quad
	\begin{subfigure}{0.475\textwidth}
		\includegraphics[width=\linewidth]{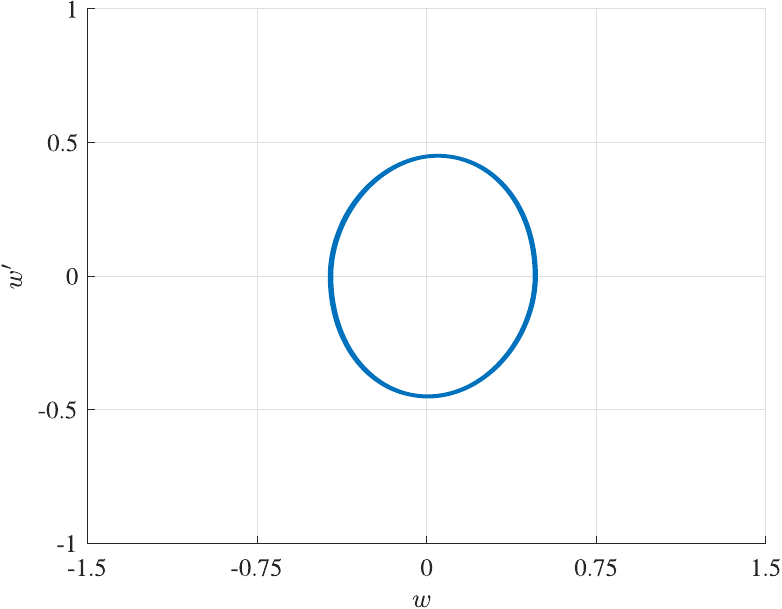}
		\caption{Phase portrait.}
		\label{Fig: sliding_absorber_PP}	
	\end{subfigure}\\
	\begin{subfigure}{0.475\textwidth}
		\includegraphics[width=\linewidth]{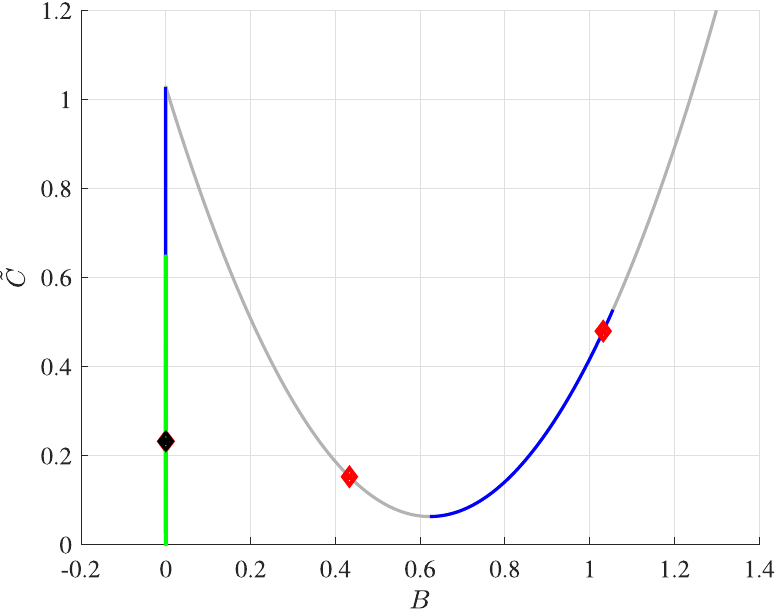}
		\caption{Projection of the flow (green line) on the $(B,\tilde{C})$-plane. The red points correspond to the equilibria of the systems from (\ref{eq: slow dynamic impacting}) and (\ref{eq: slow dynamic sliding absorber}). The black points represents the simulated steady-state amplitude.}
		\label{Fig: sliding_absorber_flow_C_B}
	\end{subfigure}\quad
	\begin{subfigure}{0.475\textwidth}
		\includegraphics[width=\linewidth]{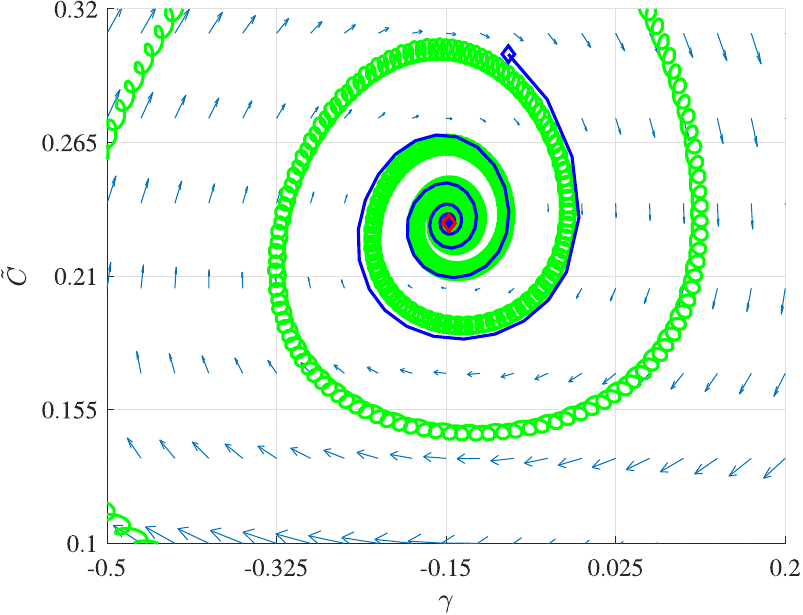}
		\caption{Projection of the flow (green line) on the $(\gamma,\tilde{C})$-plane. The blue line represents the flow of the system governed by (\ref{eq: slow dynamic sliding absorber}) and the red point indicates the corresponding equilibrium.}
		\label{Fig: sliding_absorber_flow_C_gamma}	
	\end{subfigure}
	\caption{System response and phase portraits of the absorber's periodic motion for $r=0.76$, $\mu=0.5$, $\tilde{g}=0.211$ and the external forcing parameters $G=2.47$ and $\sigma=-2.58$,  simulated using (\ref{eq: eq of motion 2a})–(\ref{eq: eq of motion 2b}). For the initial conditions $q_1(0)=q_2(0)=0 \left[\text{m}\right]$ and $\dot{q}_1(0)=\dot{q}_2(0)=0 \left[\text{m/s}\right]$, the absorber is purely sliding.}
	\label{Fig: sliding absorber}
\end{figure}
As shown in Figure~(\ref{Fig: sliding_absorber_flow_C_B}), the solution converges to the nearest stable equilibrium, located on the vertical manifold. The black point, positioned between these extreme cases, represents the steady-state equilibrium attained by the system, confirming the theoretical predictions.
The phase portrait in Figure (\ref{Fig: sliding_absorber_PP}) reflects the periodic motion of the absorber, confirming its purely sliding behavior for the given initial conditions.  Once the flow is attracted to the manifold, it remains confined to it, oscillating around the equilibrium before ultimately converging, as seen in Figure (\ref{Fig: sliding_absorber_flow_C_gamma}). 
One main difference from the frictionless VI-NES system is that, even when there are no equilibria on the SIM (\ref{eq: SIM}), the flow still potentially converges toward a steady state on the second vertical manifold. This convergence can occur after a finite number of SMR cycles. If this manifold did not exist, i.e., $\mu=0$, the strongly modulated response pattern would persist indefinitely.\\
For the second scenario, the system's flow is initialized near the SIM (\ref{eq: SIM}) of two symmetric impacts per period, with $q_1(0) = q_2(0) = 0.02\left[\text{m}\right]$ and $\dot{q}_1(0) = \dot{q}_2(0) = 0\left[\text{m/s}\right]$. 
The simulation results, presented in Figure (\ref{Fig: impacting absorber}), show that the flow converges to the stable equilibrium on the attractive branch of the SIM, represented by the black point, resulting in a steady-state response with two impacts per period. The phase portrait in Figure (\ref{Fig: impacting_absorber_PP}) illustrates the absorber's impacting behavior within the cavity, while Figures (\ref{Fig: impacting_absorber_flow_C_B}) and (\ref{Fig: impacting_absorber_flow_C_gamma}) depict the projected motion in the $(B, \tilde{C})$-plane and $(\gamma, \tilde{C})$-plane, respectively.
Together with the results from the previous section, this confirms the possible coexistence of two locally stable equilibria on both manifolds for the same external excitation. Friction enables stable periodic solutions corresponding to a purely sliding absorber, with solution branches near the stable branches of symmetric solutions with two impacts per period. These branches remain close to the linear response and suppress strongly modulated responses that would otherwise arise in the frictionless case. This coexistence is also evident in the system’s response under forward and backward frequency sweeps through resonance. This will be further examined in the next section.

\subsection{Approximation of the frequency response curves}
\begin{figure}[t!]
	\centering
	\begin{minipage}{0.575\textwidth}
		\includegraphics[width=\linewidth]{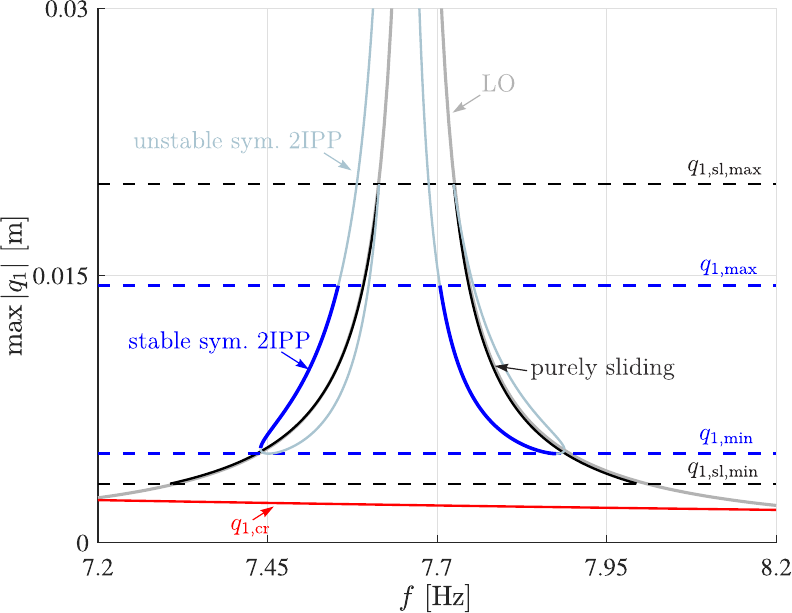}
	\end{minipage}
	\caption{Approximated frequency response curves for $r=0.76$, $\mu=0.5$ and $\tilde{g}=0.211$ and the external level of excitation $G=2.47$.}	
	\label{Fig: FRC approximation stability}	
\end{figure}

\begin{figure}[t!]
	\centering
	\begin{minipage}{0.875\textwidth}
		\includegraphics[width=\linewidth]{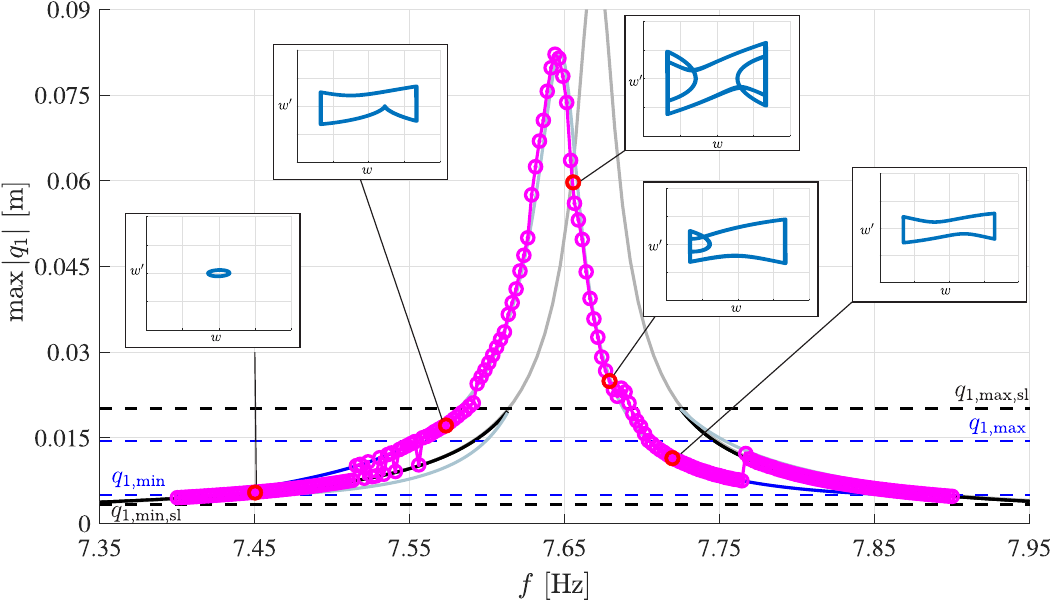}
	\end{minipage}
	\caption{System response to backward frequency sweep through resonance for $r=0.76$, $\mu=0.5$ and $\tilde{g}=0.211$ and the external level of excitation $G=2.47$.}	
	\label{Fig: FRC verification_sweep_down}	
\end{figure}

This section combines all of the analytical results to demonstrate the successful prediction of the system dynamics.  These results provide a basis for determining the optimal operating ranges of the absorber, facilitating its tuning and design. All of the above results are summarized in Figure~(\ref{Fig: FRC approximation stability}), presenting the approximated response level of the main structure, here defined as $|q_1|_{\max}$ over the excitation frequency $f$ for a fixed excitation amplitude $E$.
The derivation of the response branches and their stability assessment follow from the previous analyses. The critical amplitude threshold $q_{1,\text{cr}}$, defined in (\ref{eq: sticking limit}), is also depicted, marking the boundary between sticking and sliding motion. Three distinct periodic steady-state response regimes are identified:
\begin{enumerate}[label=(\roman*)]
	\item nonlinear response with an impacting absorber (represented by the blue lines),
	\item nonlinear response with a purely sliding absorber (represented by the black line), and 
	\item linear response with a fully sticking absorber (represented by the dashed line), which occurs only for amplitudes below the critical value $q_{1, \text{cr}}$.
\end{enumerate}
The corresponding stability bounds and bifurcation points are determined according to (\ref{eq: Cbounds}) and (\ref{eq: bounds purely sliding absorber}), and are also indicated in Figure (\ref{Fig: FRC approximation stability}). The coexistence of stable solution branches is clearly observed and is validated through time-domain simulations.
Specifically, the brute force continuation method is used to trace the system’s solution for fixed parameters at a constant excitation level $E$, while varying the input frequency $f$. Identifying all possible solutions and their stability can be computationally expensive and highly sensitive to the step size $\Delta f$. While smaller steps improve accuracy, they also increase computational cost. 
For the presented results, the external excitation $e(t)$ is defined as a harmonic base excitation with a linearly and gradually changing frequency over the frequency range of interest.
The system is then simulated for each value of the frequencies long enough for the trajectory to converge to a steady state, starting from an initial guess or previously converged trajectory. 
The simulated response level over the studied frequency range is then compared to the analytically determined branches.
\begin{figure}[t!]
	\centering
	\begin{minipage}{0.875\textwidth}
		\includegraphics[width=\linewidth]{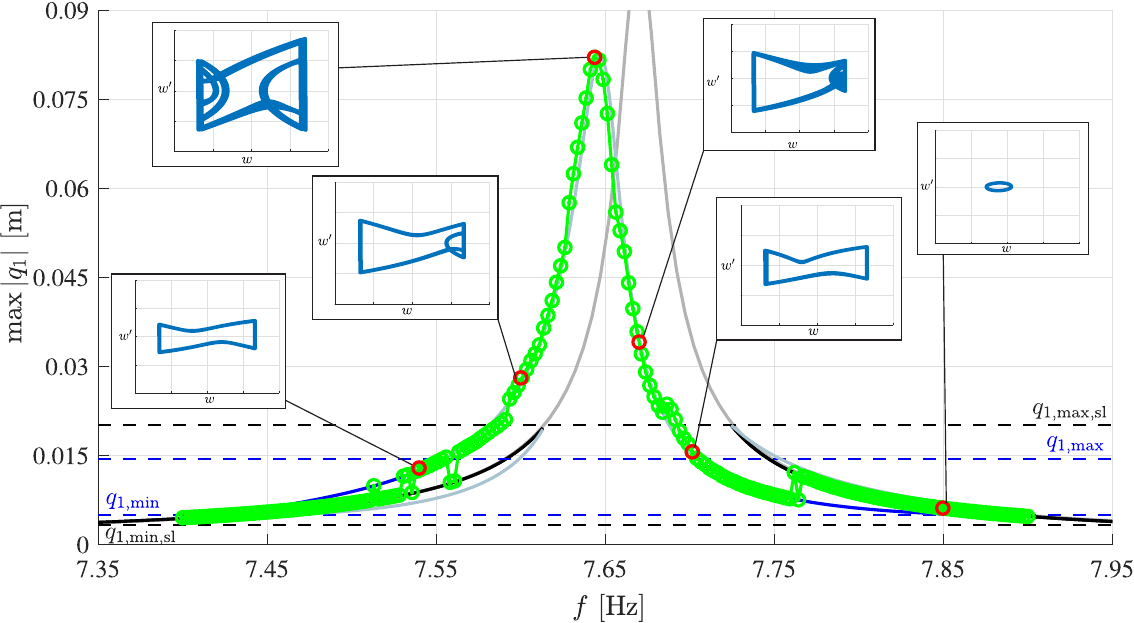}
	\end{minipage}
	\caption{System response to forward frequency sweep through resonance for $r=0.76$, $\mu=0.5$ and $\tilde{g}=0.211$ and the external level of excitation $G=2.47$.}	
	\label{Fig: FRC verification_sweep_up}	
\end{figure}
The results from forward and backward frequency sweeps, shown in Figures (\ref{Fig: FRC verification_sweep_up})-(\ref{Fig: FRC verification_sweep_down}), confirm that the system's response follows stable solution branches, indicating stable periodic regimes. Near resonance, although symmetric $\mathcal{P}_2^1$-orbits become unstable, the system transitions to other periodic solutions, such as asymmetric $\mathcal{P}_2^1$, $\mathcal{P}_3^1$, and $\mathcal{P}_4^1$-orbits, among others. Figures (\ref{Fig: FRC verification_sweep_up})-(\ref{Fig: FRC verification_sweep_down}) include phase portraits of the absorber motion during these response regimes.
This behavior aligns with the findings in Section~\ref{section: Impact Map}, confirming a symmetry-breaking bifurcation followed by a period-doubling cascade around the unstable branch of the SIM of symmetric $\mathcal{P}_2^1$-orbits. Moreover, it demonstrates the existence of additional periodic motions beyond the period-doubling sequence of $\mathcal{P}_2^1$-orbits. As the amplitude increases, the solution progresses along stable branches interconnected through symmetry-breaking and period-doubling bifurcations, forming continuous segments of the frequency response curve.
However, this continuity is not always maintained. Disruptions may occur due to the emergence of chaotic windows at the end of a period-doubling cascade or the occurrence of a fold bifurcation, where new periodic orbits appear. These phenomena can lead to gaps in the FRC, particularly near resonance, where multiple response regimes coexist.
It should be noted that, in the frequency sweep results, the observed jumps between stable branches occur either due to a loss of stability at bifurcation points or due to the chosen frequency step size, with the latter requiring finer tuning to enhance the accuracy of the simulations. 
Moreover, the presence of stable solutions corresponding to purely sliding motion of the VI-NES eliminates the strongly modulated responses that typically arise just before VI-NES activation in the frictionless case (i.e., near the stable branches of periodic solutions with two impacts per period).
In summary, the brute force continuation method has been effective in verifying the results, providing a reliable numerical approach for tracing solution paths and capturing qualitative changes in the system behavior.

\section{Conclusion} \label{section: Conlusion}
This study presents a comprehensive analytical and numerical investigation of the nonlinear dynamics of a symmetric vibro-impact nonlinear energy sink (VI-NES) with dry friction, extending previously idealized frictionless models toward more realistic configurations by accounting for dry friction. The analysis combines the Multiple Scales Method (MSM) and the impact-map approach within a unified framework that captures both the slow and fast time-scale dynamics of frictional vibro-impact systems.\\
The first key contribution is the extension of the MSM to include friction, enabling the derivation of closed-form expressions for the slow invariant manifold of the two-impacts-per-period ($2$-IPP) regime and a first-order approximation of the system’s slow dynamics on it. This inclusion broadens the applicability of the MSM and provides deeper insight into the nonlinear near-resonant behavior of the system. It also yields closed-form expressions for design-relevant performance metrics, such as the VI-NES activation threshold, effective dissipation characteristics, and backbone curve. These metrics define the engagement conditions and energy-transfer efficiency of the absorber during targeted energy transfer.\\
The complementary impact-map approach captures the fast evolution of trajectories through discrete impact events and provides closed-form expressions for the amplitude bounds and stability limits of the $2$-IPP regime. Together, these analytical results define the optimal operating range of the VI-NES and establish a consistent foundation for parameter tuning and design optimization. The strong agreement between both methods reinforces the validity of the multiple scales analysis and demonstrates the robustness of the impact-map approach in predicting periodic responses, their stability, and bifurcations.\\
Another key contribution is the use of the extended MSM formulation to investigate the coexistence of response regimes near the optimal operating range of the VI-NES. The analysis reveals two distinct responses at the same amplitude levels: one dominated by impacts and the other governed by continuous sliding. Numerical results validate these analytical predictions and the computed branches of periodic solutions, confirming the transitions between regimes as excitation conditions vary. The study also demonstrates that friction introduces competing effects that depend strongly on vibration amplitude. At low amplitudes, friction enhances dissipation and raises the activation threshold, improving energy-transfer efficiency but reducing robustness. At higher amplitudes, however, its influence diminishes, and the system behavior converges toward the idealized frictionless case. This dual role emphasizes the need to balance robustness and efficiency when tuning VI-NES parameters for optimal performance.\\
This work establishes a rigorous framework for understanding the dynamics of VI-NES subjected to friction and provides analytical and numerical tools for system design and optimization. It also forms the theoretical foundation for a forthcoming experimental study that will validate the analytical results and assess the VI-NES performance under real excitation conditions.

\section*{Acknowledgement}
This research was supported by the Deutsche Forschungsgemeinschaft, DFG, project number 402813361.

\section*{Competing Interests}
The authors declare that they have no competing interests.

\bibliographystyle{unsrt} 
\bibliography{references}

\end{document}